\documentclass[referee]{aa} 
\usepackage{graphicx}
\usepackage{txfonts}
\usepackage{longtable}
\usepackage{lscape}

\usepackage{natbib}
\bibpunct{(}{)}{;}{a}{}{,} 

\usepackage{colortbl}
\definecolor{gray}{rgb}{0.6941,0.6352941176470588,0.6352941176470588}

\newif\ifshowImage\showImagetrue 

\newif\ifshowRecord

\newcommand{\dotcirc}{$_{^\centerdot}^\circ$}

\newcommand{\imagesPath}{.}

\newcommand{\makeFigureCOComponentes}{
\begin{figure*}[!th]
\ifshowImage

\resizebox{\textwidth}{!}{\includegraphics{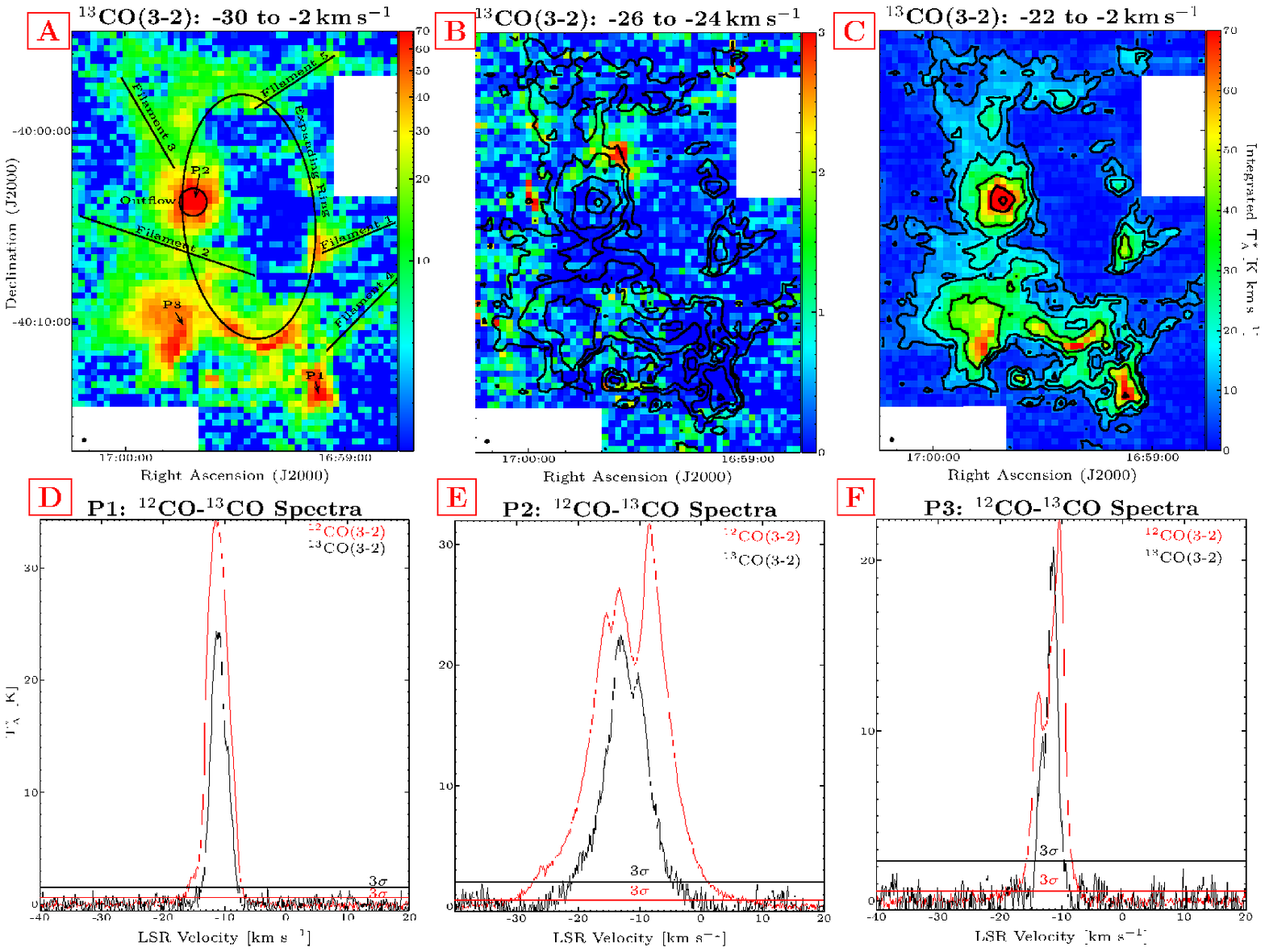}}
\fi
\caption{
Top panels: integrated $^{13}$CO(3-2) line of the ring G345.50+1.50 between
$-$30 and $-$2\,km\,s$^{-1}$ (panel a), between $-$26 and
$-$24\,km\,s$^{-1}$ (panel b), and between $-22$ and
$-$2\,km\,s$^{-1}$ (panel c).  The black circles in the bottom left corner
represent the APEX beamsize at 330.588\,GHz, $\sim$18$''$.  Contours
show the emission integrated between $-$30 and $-$2\,km\,s$^{-1}$ with
levels of 9, 18, 36, 72, and 144\,K\,km\,s$^{-1}$.  In panel a, the
ellipse indicates the expanding ring model in the X-Y plane, the lines
show the filaments, and the circle plots the outflow IRAS\,16562-3959.
Panels d, e, and f show the $^{12}$CO(3-2) and $^{13}$CO(3-2) spectra
toward the positions P1, P2, and P3, respectively, indicated in panel
a with the arrows.
}
\label{figure13COComponentes}
\end{figure*}
}

\newcommand{\makeFigureMGPSMSXSPITZER}{
\begin{figure*}[!th]
\ifshowImage
\resizebox{\hsize}{!}{\includegraphics{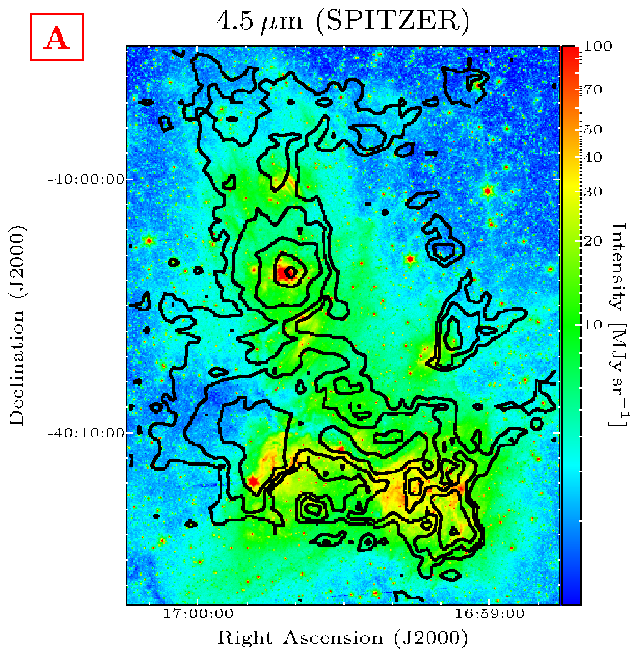}\includegraphics{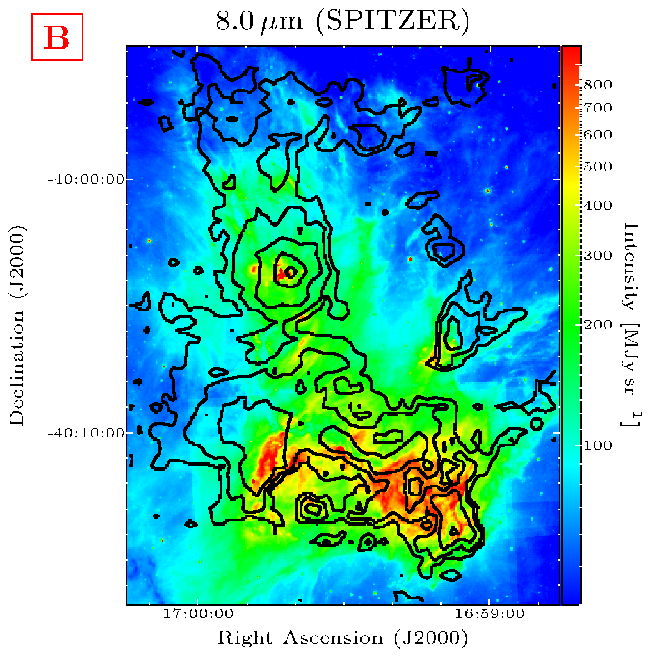}}
\resizebox{\hsize}{!}{\includegraphics{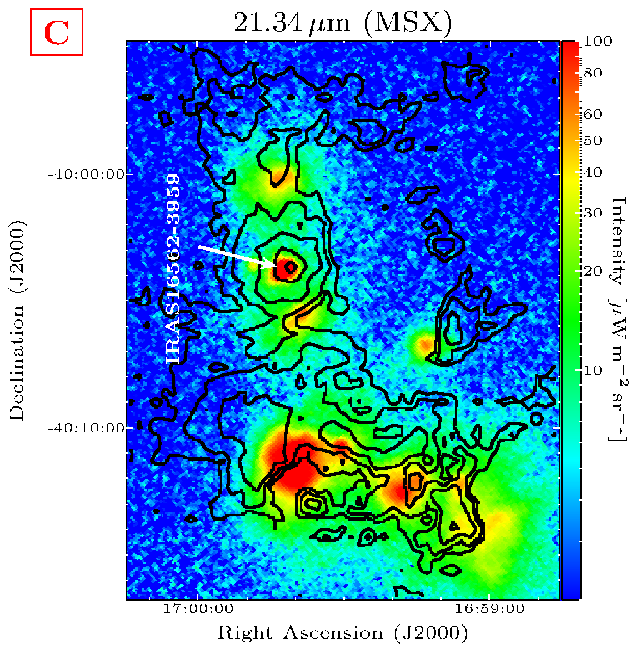}\includegraphics{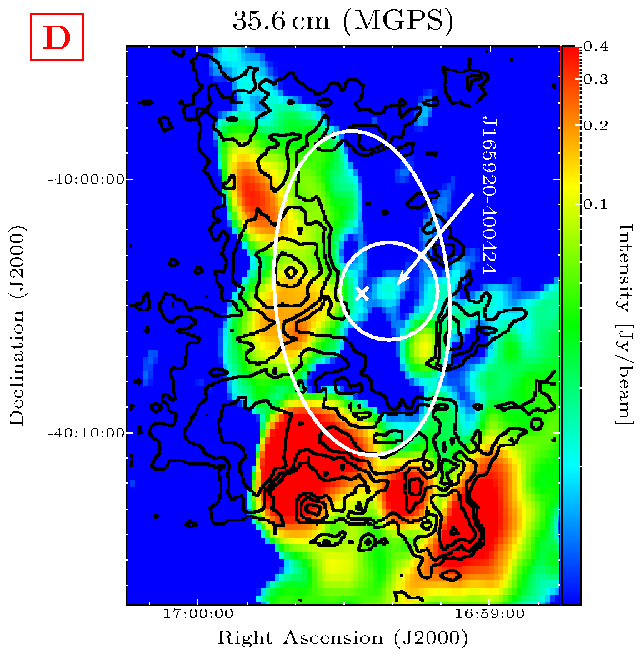}}
\fi
\caption{
Images of radio and infrared continuum emission toward the ring
G345.50+1.50, overlaid with contours of the $^{13}$CO(3-2) line
integrated between $-$30 and $-$2\,km\,s$^{-1}$ (levels: 9, 18, 36,
72, and 144\,K\,km\,s$^{-1}$).  Panels a and b: 4.5 and 8.0 \,$\mu$m
images from Spitzer.  Panel c: 21.34\,$\mu$m image from MSX.  Panel d:
35.6\,cm image from MGPS \citep{murphy2007}. The arrow indicates the
position of the 35.6\,cm source J165920-400424, and the 2$'$ white
circle shows the area where objects were searched for in the NED
database to determine the nature of this source.  The ellipse
indicates the expanding ring model in the X-Y plane, with its center
indicated with an X, see Sect. \ref{sectionExpandingRing}.
}
\label{figureMGPS-MSX-SPITZER}
\end{figure*}
}

\newcommand{\makeFigureDeltaN}{
\begin{figure}[!th]
\ifshowImage
\resizebox{\hsize}{!}{\includegraphics{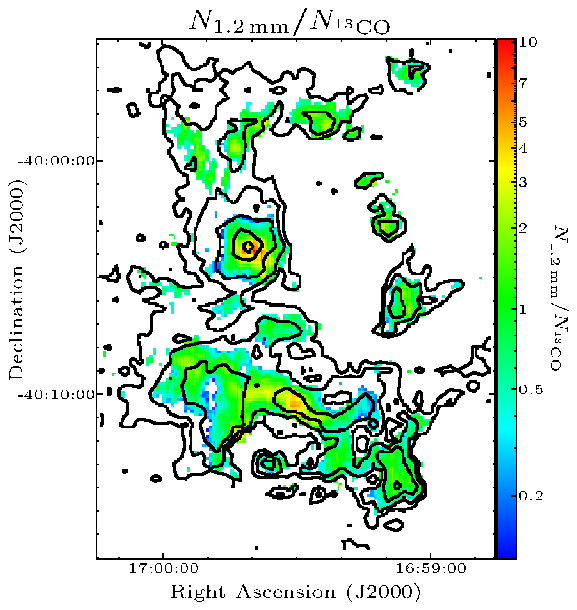}}
\fi
\caption{
Map of the column density ratio
$N_{\mathrm{1.2mm}}$/$N_{\mathrm{^{13}CO}}$ with contours of the
$^{13}$CO(3-2) line integrated between $-$30 and $-$2\,km\,s$^{-1}$
(levels: 9, 18, 36, 72, and 144\,K\,km\,s$^{-1}$).  Only significant
column density values, higher than $3\sigma$, have been used in the
ratio map.
}
\label{figureDeltaN}
\end{figure}}

\newcommand{\makeTableSummary}{
\begin{table}
\caption{Summary of the physical properties of the 104 identified
    $^{13}$CO clumps.}  
\label{tableSummary}
\begin{center}
\begin{tabular}{lccc}
\hline\hline\\
Parameter            & Range      & Average & Median \\
\hline\\
Diameter & 0.3-1.6&    0.9 &   0.9 \\
$[$pc]           & &    &    \\
$^1$Mass & 2.3-7.5$\times$10$^2$ & 44  & 23 \\
$[$M$_\odot$]   & & & \\
$^2$Density  & 10$^2$-10$^4$&  2$\times$10$^3$& 10$^3$\\
$[$cm$^{-3}$]& &  & \\
$^2$Column density& 3$\times$10$^{20}$-2$\times$10$^{22}$ & 3$\times$10$^{21}$&2$\times$10$^{21}$\\
 $[$cm$^{-2}$]&&&\\
 Line velocity width& 0.7-2.0&1.1&1.0\\
$[$km\,s$^{-1}$]&&&\\
\hline
\end{tabular}
\end{center}
\begin{list}{}{}
 \item[$^1$]The total mass of the clumps is 4.5$\times$10$^3$\,M$_\odot$.
 \item[$^2$]Densities and column densities are estimated assuming a mean molecular weight of $\mu$=2.29.
\end{list}{}{}
\end{table}
}

\begin{document}
\title{G345.45+1.50: An expanding ring-like structure with massive star formation
\thanks{ This publication is based on data acquired with the Atacama Pathfinder Experiment (APEX). APEX  is a collaboration between the Max-Planck Institut f\"ur Radioastronomie, the European Southern Observatory, and the Onsala Space Observatory.}
\fnmsep\thanks{ The Atacama Submillimeter Telescope (ASTE) Experiment is a project driven by the National Astronomical Observatory of Japan in collaboration with Universidad de Chile, and Japanese institutes including University of Tokyo, Nagoya University, Osaka-Prefecture University, Ibaragi University, and Hokkaido University.}
}


   \author{ Cristian L\'opez-Calder\'on
          \inst{1,2,3}
          \and
          Leonardo Bronfman
          \inst{1}
          \and
          Lars-{\AA}ke Nyman
          \inst{2,4}
          \and
          Guido Garay
          \inst{1}
          \and
          Itziar de Gregorio-Monsalvo
          \inst{2,4,5}
          \and
          Per Bergman
          \inst{6}
          }

   \institute{Departamento de Astronom\'{\i}a, Universidad de Chile,
              Casilla 36-D, Santiago, Chile.
              \email{clopez@das.uchile.cl}
           \and
           { Joint ALMA Observatory (JAO), Alonso de C\'ordova 3107, Vitacura, Santiago, Chile}
           \and
           National Radio Astronomy Observatory, Charlottesville, VA 22903, USA
           \and
           { European Southern Observatory, Alonso de C\'ordova 3107, Vitacura, Santiago, Chile}
           \and
           { ESO Garching, Karl-Schwarzschild Str. 2, 85748 Garching, Germany}
           \and
           Onsala Space Observatory, Chalmers Univ. of Technology, 439 92 Onsala, Sweden
   }

   \date{}

\newcommand{\circdot}{_{^\centerdot}^\circ}

\abstract{ 
Ring-like structures in the interstellar medium (ISM) are commonly
associated with high-mass stars.  Kinematic studies of large
structures in giant molecular clouds (GMCs) toward these ring-like
structures may help us to understand how massive stars form.
}{
The origin and properties of the \object{ring-like structure G345.45+1.50} is
investigated through observations of the $^{13}$CO(3-2) line.  The aim
of the observations is to determine the kinematics in the region and
to compare physical characteristics estimated from gas emission with
those previously determined using dust continuum emission.  This area
in the sky is well suited for studies like this because the ring is
located 1$\circdot$5 above the Galactic plane at 1.8\,kpc from the
Sun, thus molecular structures are rarely superposed on our line of
sight, which minimizes confusion effects that might hinder identifying
of individual molecular condensations.
}{
The $^{13}$CO(3-2) line was mapped toward the whole ring using the
Atacama Pathfinder Experiment (APEX) telescope.  The observations
cover 17$'$$\times$20$'$ in the sky with a spatial resolution of
0.2\,pc and an rms of $\sim$1\,K at a spectral resolution of
0.1\,km\,s$^{-1}$.
}{
The ring is found to be expanding with a velocity of
1.0\,km\,s$^{-1}$, containing a total mass of
6.9$\times$10$^3$\,M$_\odot$, which agrees well with that determined
using 1.2\,mm dust continuum emission.  An expansion timescale of
$\sim$3$\times$10$^6$\,yr and a total energy of
$\sim$7$\times$10$^{46}$\,erg are estimated.  The origin of the ring
might have been a supernova explosion, since a 35.5\,cm source,
J165920-400424, is located at the center of the ring without an
infrared counterpart.  The ring is fragmented, and 104 clumps were
identified with diameters of between 0.3 and 1.6\,pc, masses of
between 2.3 and 7.5$\times$10$^2$\,M$_\odot$, and densities of between
$\sim$10$^{2}$ and $\sim$10$^{4}$\,cm$^{-3}$.  At least 18\% of the
clumps are forming stars, as is shown in infrared images.  Assuming
that the clumps can be modeled as Bonnor-Ebert spheres, 13 clumps are
collapsing, and the rest of them are in hydrostatic equilibrium with
an external pressure with a median value of
4$\times$10$^4$\,K\,cm$^{-3}$.  In the region, the molecular outflow
IRAS\,16562-3959 is identified, with a velocity range of
38.4~km~s$^{-1}$, total mass of 13~M$_\odot$, and kinematic energy of
7$\times$10$^{45}$~erg.  Finally, five filamentary structures were
found at the edge of the ring with an average size of 3\,pc, a width
of 0.6\,pc, a mass of 2$\times$10$^2$~M$_\odot$, and a column density
of 6$\times$10$^{21}$\,cm$^{-2}$.
}{}

\keywords{ISM: clouds, dust, molecules, kinematics and dynamics -- stars: massive,  formation}

\maketitle

\section{Introduction}

A wealth of observations have shown that ring- and shell-like
structures in the interstellar medium (ISM) are ubiquitous
\citep[e.g.][]{churchwell2006,churchwell2007,churchwell2008,heiles1979,martin-pintado1999,oka1998,schuller2009,wong2008,beaumont2010,sidorin2014,anderson2015}.
In the Galaxy, these structures range up to 1.2\,kpc in radius,
2x10$^7$ M$_\odot$ in mass, and 10$^{53}$\,erg in kinetic energy.
Their formation has been associated with the energy released by
massive stars through stellar winds, supernova explosions, and
ionizing radiation, or by collisions of high-velocity HI clouds with
the Galactic disk \citep{tenorio-tagle1988}.  Rings with a size
smaller than 100\,pc have been related with OB associations and
stellar clusters in their interiors \citep{tenorio-tagle1988}, and hot
cores and molecular clumps at their edges
\citep[e.g.][]{martin-pintado1999,wong2008}.

In this study, the entire ring-like structure G345.45+1.50 was
observed in the $^{13}$CO(3-2) line to study its properties including
the kinematics.  The ring is located $\sim$1$\circdot$5 above the
Galactic plane, so there is little contamination with foreground and
background molecular structures, which are mainly concentrated toward
the Galactic plane, thus minimizing confusion effects in identifying
individual condensations.  Properties of the ring and of the
identified clumps are estimated from $^{13}$CO(3-2) line observations,
and compared with those found using the 1.2\,mm continuum emission.
The molecular structure is also compared with infrared observations,
from MSX\footnote{Downloadable from http://irsa.ipac.caltech.edu} and
Spitzer$^1$.

This ring is located at a distance of 1.8 kpc from the Sun, inside the
\object{GMC G345.5+1.0}.  This GMC is located approximately between
344\dotcirc5 and 346\dotcirc5 in Galactic longitude, between
0\dotcirc2 and 2\dotcirc0 in Galactic latitude, and between $-$33 and
$-$2 km\,s$^{-1}$ in LSR velocity \citep{bronfman1989}.  The ring-like
structure contains a total mass of $\sim$4.0$\times$10$^3$\,M$_\odot$
estimated through 1.2\,mm continuum emission \citep{lopez2011}.  The
1.2\,mm emission is fragmented at a spatial resolution of
$\sim$0.2\,pc and composed of $\sim$54 clumps, which have an average
mass of 75\,M$_\odot$.  Nineteen clumps are associated with infrared
sources identified in the Midcourse Space Experiment
\citep[MSX;][]{price2001} and Spitzer observations
\citep{benjamin2003}, including the source \object{IRAS\,16562-3959}, a massive
star-forming region (MSFR) with a luminosity of
$\sim$5.3$\times$10$^4$\,L$_\odot$ and associated with the most
massive and dense clump in the ring.  Within the ring lie also 35 cold
clumps, that is, dense condensations that are not associated with an
infrared counterpart.  They might be regarded as candidates in which
massive star formation has not yet started.


\begin{table}
\caption{List of observations with the ASTE telescope of the
  $^{12}$CO(3-2) line.  Columns 1 and 2 show equatorial coordinates,
  Col. 3 the antenna temperature intensity peaks, and Col. 4
  the estimated kinetic temperatures.}
\begin{center}
\label{tableASTE}
\begin{tabular}{cccc}
\hline\hline\\
R.A. & Dec.&  $^1$$T^*_\mathrm{A}$ & $^1$$T_\mathrm{K}$  \\     
\multicolumn{2}{c}{J2000}         & K        &  K      \\
\hline\hline\\

16:59:08&-40:13:57&34&57\\
16:59:41&-40:03:37&32&53\\
16:59:44&-40:10:17&23&40\\
16:59:46&-40:11:37&25&43\\
16:59:20&-40:11:17&20&36\\
16:59:13&-40:11:17&23&41\\
16:59:08&-40:05:37&19&35\\
16:59:32&-40:10:17&15&29\\
16:59:35&-40:07:37&26&45\\
16:59:04&-40:11:17&20&36\\
16:59:23&-40:13:17&16&30\\
16:59:35&-40:12:57&21&38\\
16:59:53&-40:08:37&9.7&21\\
16:59:20&-40:09:37&14&28\\
16:59:11&-40:02:37&14&27\\
16:59:28&-39:57:57&13&25\\
16:59:23&-39:58:37&14&27\\
16:59:51&-40:00:37&18&34\\
16:59:44&-39:59:17&12&25\\
16:59:32&-39:58:17&13&26\\
16:59:16&-40:00:57&15&29\\
16:59:18&-39:58:57&13&26\\
16:59:30&-39:56:37&9.6&21\\
16:59:06&-39:55:57&11&23\\
16:58:50&-40:08:16&13&26\\
16:59:56&-39:58:37&11&23\\
16:59:49&-39:56:17&7.2&17\\
16:58:52&-40:09:16&15&29\\
16:59:58&-40:05:57&7.7&18\\

\hline
\end{tabular}\\
 $^1$ The error for $T_\mathrm{A}^*$ and $T_\mathrm{K}$ values is $\sim$0.2\,K.
\end{center}

\end{table}

\section{Observations}
\label{sectionObservations}
\subsection{$^{13}$CO(3-2) line}

The $^{13}$CO(3-2) line observations were made at 330.588\,GHz using
on-the-fly (OTF) observing mode with the APEX-2A heterodyne receiver
mounted at the APEX telescope \citep{gusten2006,risacher2006}.  The
whole ring-like structure G345.45+1.50 was mapped using a total time
on source of $\sim$14\,hr on October 17 - 19, 2005, May 28, June 24 -
26, and October 7 - 10, 2006. The steps in the map were 6\,arcsec and
the dump integration time was 1.0\,sec.  At 331\,GHz the telescope has
a beam size of 18$''$, corresponding to $\sim$0.2\,pc at 1.8\,kpc from
the Sun, and a main-beam efficiency of $\sim$0.73. The spectrometer
had a bandwidth of 1\,GHz with 8192 channels, giving a spectral
resolution of 122\,kHz, corresponding to $\sim$0.1\,km\,s$^{-1}$ at
this frequency.

The observations were reduced using
XS\footnote{http://www.chalmers.se/rss/oso-en/observations/data-reduction-software}
software, resulting in a data cube with a projected size of
$\sim$17$\times$22\,arcmin$^2$ and a pixel size of 20$''$, consisting
of 3036 spectra with an average rms of $\sim$1\,K (antenna
temperature) at a spectral resolution of $\sim$0.1\,km\,s$^{-1}$.

\begin{figure}[!th]
\ifshowImage
\resizebox{\hsize}{!}{\includegraphics{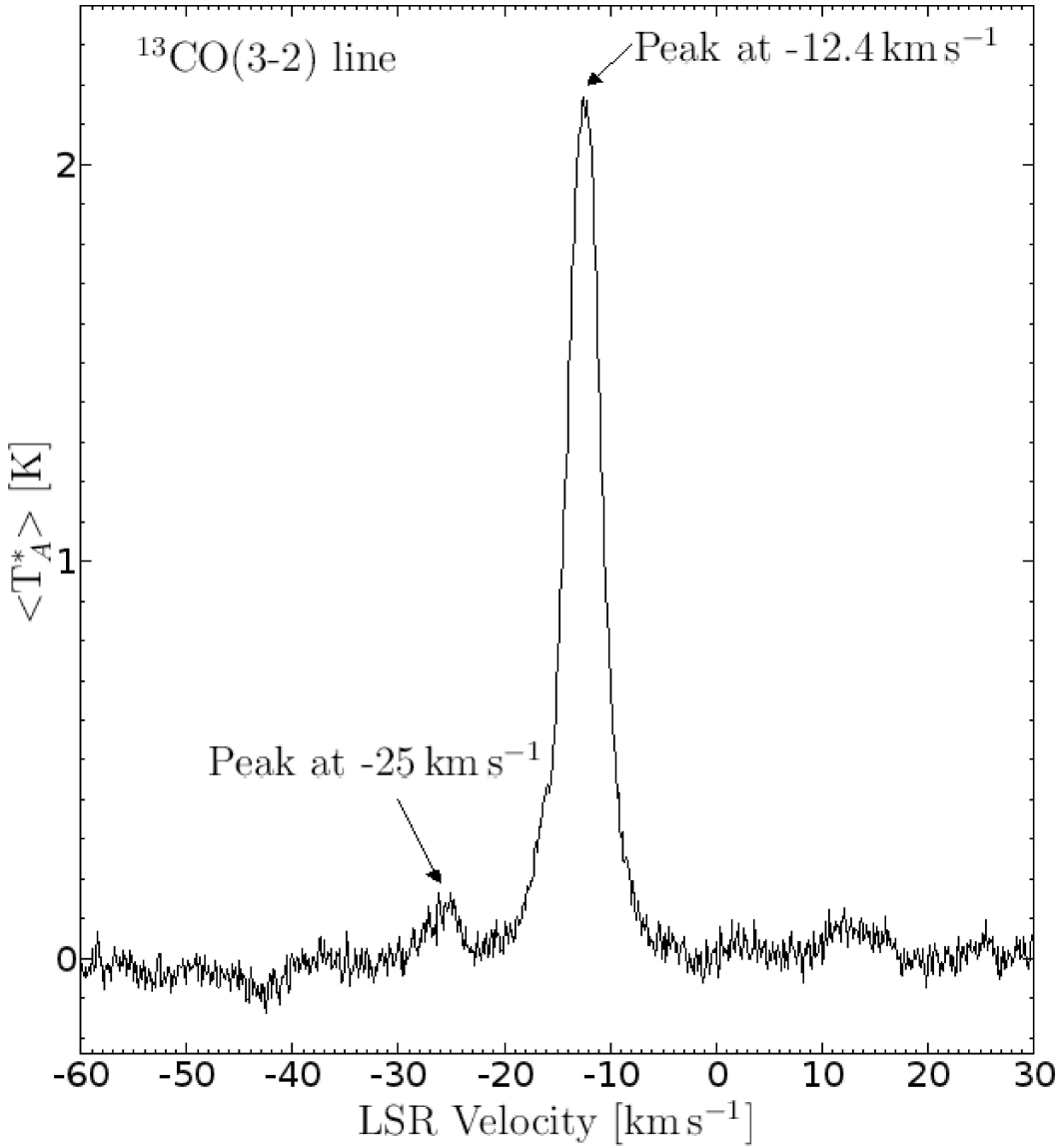}}
\fi
\caption{Average of all $^{13}$CO(3-2) spectra observed toward the
  ring-like structure G345.45+1.50.  $<T_\mathrm{A}^*>$ denotes the
  antenna temperature average.  Arrows indicate the two peaks in the
  emission.}
\label{figure13COspectrum}
\end{figure}

\makeFigureCOComponentes

\subsection{$^{12}$CO(3-2) line}

To estimate the gas temperature in the region, the $^{12}$CO(3-2) line
was observed using ASTE toward 29 intensity peaks found in the
$^{13}$CO(3-2) line map.  Observations were made on August 27 2010
using the position-switching observing mode.  Table \ref{tableASTE}
lists the positions of the observed peaks.  The CATS-345\,GHz receiver
was used, with on-source integration times of between 20 and 30
seconds.  The bandwidth of the spectrometer was 128\,MHz with 1024
channels, providing a spectral resolution of 125\,kHz, or
$\sim$0.1\,km\,s$^{-1}$ at the $^{12}$CO(3-2) line transition
frequency, $\sim$345.796\,GHz.  At this frequency, the beam size of
the ASTE telescope is $\sim$22$''$, $\sim$0.2\,pc at 1.8\,kpc from the
Sun.  The estimated main-beam efficiency of the telescope is
$\sim$0.7.

The data were reduced using the
NEWSTAR\footnote{www.nro.nao.ac.jp/$\sim$nro45mrt/obs/newstar}
software, resulting in 29 spectra with an antenna temperature rms of
$\sim$0.2\,K at a spectral resolution of 0.1\,km\,s$^{-1}$.

\section{Results and discussion}
\label{sectionResults}

\subsection{$^{13}$CO(3-2) line}

The $^{13}$CO(3-2) line toward the ring has velocities between $-30$
and $-$2\,km\,s$^{-1}$, with two gas components; the stronger peaks at
$-12.4$\,km\,s$^{-1}$ and the weaker at $-25$\,km\,s$^{-1}$.  Figure
\ref{figure13COspectrum} shows the globally averaged spectrum of the
$^{13}$CO(3-2) line over the whole ring.  The velocity range agrees
with the hypothesis that the ring is part of GMC G345.5+1.0, which has
velocity components of between $-33$ and $-2$\,km\,s$^{-1}$.  The two
velocity components observed in the $^{13}$CO(3-2) line were
previously identified in the $^{12}$CO(1-0) line \citep{bronfman1989},
and the stronger component in the CS(2-1) line \citep{bronfman1996}.

%
%

The spatial distributions of the $^{13}$CO(3-2) line for the whole
ring and for the two velocity components are shown in
Figs. \ref{figure13COComponentes}a, \ref{figure13COComponentes}b, and
\ref{figure13COComponentes}c, integrated in three velocity ranges,
$-30$ to $-2$\,km\,s$^{-1}$, $-26$ to $-24$\,km\,s$^{-1}$, and $-22$
to $-$2\,km\,s$^{-1}$, respectively.  To display the weaker component,
the velocity range $-26$ to $-24$\,km\,s$^{-1}$ was used because the
resulting integrated image is less noisy than that using the velocity
range $-30$ to $-22$\,km\,s$^{-1}$.  The spatial structure of the
stronger component ($-$22 to $-$2~km\,s$^{-1}$) contains most of the
emission and is composed of several condensations that form a
ring-like shape.  Figure \ref{figure13COComponentes}c shows that the
weaker component corresponds to a small condensation.

\subsection{Star formation process along the ring}
\label{sectionStarFormationProcessAlongTheRing}

%
%

It is possible to distinguish different stages in the star formation
process along the ring by comparing the $^{13}$CO(3-2) line with
observations made at other wavelengths.

%
%

\begin{figure}
\ifshowImage
\resizebox{\hsize}{!}{\includegraphics{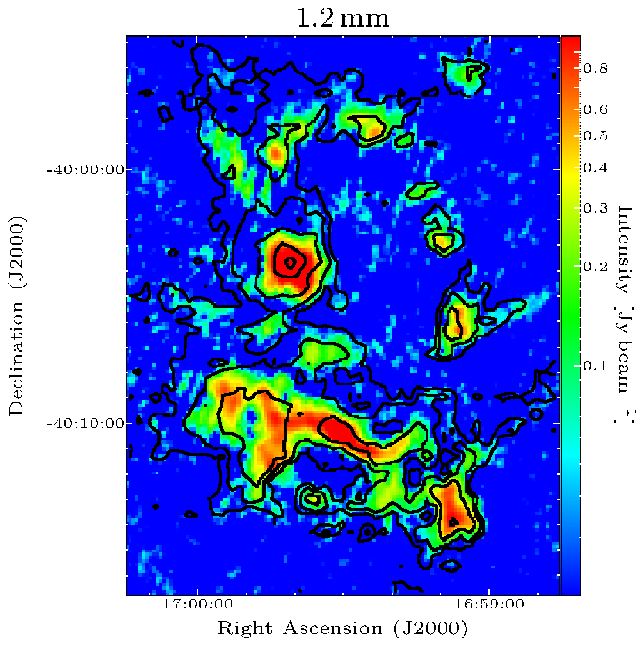}}
\fi
\caption{
1.2\,mm continuum emission of the ring G345.50+1.50 \citep{lopez2011},
with contours of the $^{13}$CO(3-2) line integrated between $-$30 and
$-$2\,km\,s$^{-1}$ (levels: 9, 18, 36, 72, and 144\,K\,km\,s$^{-1}$).
}
\label{figureSIMBARING}
\end{figure}

Figure \ref{figureSIMBARING} shows the 1.2\,mm continuum emission
overlaid with contours of the integrated $^{13}$CO line intensity.
The $^{13}$CO(3-2) line delineates the cold dust structure traced in
1.2\,mm continuum emission, exhibiting the same ring-like structure;
but the $^{13}$CO(3-2) line is more extended than the 1.2\,mm
continuum.  These cold gas and dust components enclose the emission
generated by hot dust heated by embedded stars, which are observed in
the MSX and Spitzer images displayed in
Figs. \ref{figureMGPS-MSX-SPITZER}a, \ref{figureMGPS-MSX-SPITZER}b,
and \ref{figureMGPS-MSX-SPITZER}c.

%
%

Cold condensations are the sites where stars might eventually be
formed.  These condensations are identified along the ring, being
traced by the $^{13}$CO(3-2) line or the 1.2\,mm continuum, but
without an infrared counterpart in MSX and Spitzer images, mainly in
21.34 $\mu$m MSX band, since the other infrared bands are contaminated
by polycyclic aromatic hydrocarbon (PAH) emission and photospheric
emission from stars \citep[e.g.][]{chavarria2008}.

%
%

Figure \ref{figureMGPS-MSX-SPITZER}d shows the 35.6\,cm continuum
image taken from Molonglo Galactic Plane Survey
\citep[MGPS;][]{murphy2007}, which is thought to correspond mainly to
free-free emission, tracing HII regions.  These HII regions have
counterparts at infrared wavelengths; J165920-400424, a 35.5\,cm
source from MGPS located at RA=16:59:29 Dec=$-$40:04:23 (J2000), does
not have a clear infrared counterpart, however. It might be a pulsar
produced by a supernova explosion (see
Sect. \ref{sectionExpandingRing}), but other possibilities such as an
extragalactic source, a radio star, or planetary nebulae cannot be
ruled out \citep[e.g.][]{whiteoak1992}.  To reveal the nature of this
source, the NED\footnote{http://ned.ipac.caltech.edu/} and
Vizier\footnote{http://vizier.u-strasbg.fr/viz-bin/VizieR} databases
were used.  No object was found within 2\,arcmin in NED, and only
stellar objects are found in Vizier within 25\,arcsec, approximately
half of the beam-size of the MGPS observations.  To clarify the nature
of this source, ATCA observations would be needed toward
J165920-400424 in cm wavelengths, allowing us to determine whether the
emission is synchrotron or bremsstrahlung.

%
%

Features at 7.7 and 8.6\,$\mu$m, attributed to vibrational emission of
PAHs excited by UV radiation \citep[e.g.][]{vanDishoeck2004}, should
be observed in the 8.0\,$\mu$m Spitzer band, while the 4.5\,$\mu$m
Spitzer band is expected to contain no PAH features, and therefore
traces the continuum emission \citep[][]{benjamin2003}.  The emission
in the 4.5 and 8.0\,$\mu$m bands toward the ring are roughly similar,
but the 8.0\,$\mu$m emission is more extended and delineates the
structures better, possible by emission of PAHs.

\subsection{Gas temperature}
\label{subsectionGasTemperature}

%
%

The gas kinetic temperature, $T_{\mathrm k}$, was estimated from our
$^{12}$CO(3-2) observations by assuming optically thick emission and
that the excitation temperature can be approximated by the kinetic
temperature. The peak brightness, $T_{\mathrm B}$, of the line is then
\begin{equation}
T_{\mathrm B} = \frac{h\nu}{k}\,\frac{1}{\exp\left(h\nu/kT_{\mathrm
    k}\right) - 1},
\end{equation}
where $\nu$ is the frequency, $h$ and $k$ are the Planck and Boltzmann
constants, respectively.  We recall that estimating $T_{\mathrm k}$ in
this way results in a lower limit, since the source may not fill the
the beam entirely.  Absorption taking place in lower excitation
foreground gas will also yield a lower estimate.  For example, Fig. 2d
shows a spectrum without absorption, but Figs. 2e and 2f show signs of
self-absorption in some positions.  Assuming a beam-filling factor
($f_{\mathrm{BEAM}}$) of 1, $T_{\mathrm{B}}$ can be determined from
\begin{equation}
\label{equationBrightnessTemperature}
\qquad\qquad\qquad
f_{\mathrm{BEAM}} T_{\mathrm{B}}\sim T_{\mathrm{MB}} \sim T^{\mathrm{*}}_{\mathrm{A}}/\eta_{\mathrm{MB}},
\end{equation}
where $T_{\mathrm{MB}}$ is the peak of the main-beam brightness
temperature, $T^{\mathrm{*}}_{\mathrm{A}}$ is the peak of the antenna
temperature, and $\eta_{\mathrm{MB}}$ is the main-beam efficiency,
$\sim$0.7.

%
%

$T^{\mathrm*}_{\mathrm{A}}$ varies between 7.2 and 34\,K, implying a
variation of $T_{\mathrm{K}}$ between 17 and 57\,K, with an average
value of $\sim$30\,K as is shown in Table \ref{tableASTE}, consistent
with the dust temperature of $\sim$30\,K estimated for MSFRs
\citep{faundez2004}.

%
%

\makeFigureMGPSMSXSPITZER

\subsection{Column density of the ring}
\label{sectionColumnDensityOfTheRing}

%
%

Assuming that the $^{13}$CO(3-2) line emission is optically thin and
in local thermal equilibrium (LTE), the total column density can be
determined using Eq. \ref{equationLTE} in Appendix I.  From the
integrated $^{13}$CO(3-2) line {intensity} between $-30$ and
$-2$\,km\,s$^{-1}$, the column density of the ring varies between
2$\times$10$^{21}$ and 7$\times$10$^{22}$\,cm$^{-2}$, with an average
of 7$\times$10$^{21}$\,cm$^{-2}$.

%
%

The column density of the ring can also be estimated from the 1.2\,mm
dust continuum emission, $N_\mathrm{1.2mm}$.  This emission is assumed
to be optically thin, thus the total column density is given by
Eq. \ref{equationDust} in Appendix II.

%
%

Figure \ref{figureDeltaN} shows the ratio of $N_\mathrm{1.2\,mm}$ to
$N_\mathrm{^{13}CO}$ overlaid with contours of the integrated
$^{13}$CO(3-2) line, which is proportional to the the column density
following Eq. \ref{equationDust}.  The ratio
$N_\mathrm{1.2\,mm}/N_\mathrm{^{13}CO}$ varies from $\sim$0.1 to 10
with an average value of $\sim$1.0.  $N_\mathrm{1.2\,mm}$ is higher
than $N_\mathrm{^{13}CO}$ toward the central and dense parts of the
condensations.  This is likely caused by optically thick $^{13}$CO
emission.  To investigate this in more detail, we would need C$^{18}$O
observations toward the central parts and also continuum data at
shorter wavelengths than 1.2\,mm.  Low N$_{1.2mm}$/N$^{13}$CO ratios
are generally found near the edges of the condensations where the
density is presumingly lower. The reason for this is not clear, but
$T_K$ $>$ $T_{dust}$ or different dust properties near the edges might
explain it.

%
%

The assumed values of the physical parameters can introduce variations
in the estimated ratio $N_\mathrm{1.2\,mm}/N_\mathrm{^{13}CO}$.  For
example, \cite{faundez2004} found massive star-forming regions with
dust temperatures of between 18 and 46\,K, which produce variations in
this ratio by a factor 2.  \cite{hildebrand1983} estimated a ratio of
gas to dust masses of 100 with errors within a factor 2-8.
\cite{frerking1982} estimated a [H$_2$/$^{13}$CO] of 7$\times$10$^5$
with possible variations of a factor 2-10.  \cite{ossenkopf1994}
computed a dust opacity of $\sim$1 cm$^2$/g at 1.2\,mm, with
deviations no more than a factor 2.  Beam dilution is another source
of error; when there is beam dilution, the ratio
$N_\mathrm{1.2\,mm}/N_\mathrm{^{13}CO}$ has to be corrected by a
factor
$\Omega_\mathrm{MB-SEST}^2/\Omega_\mathrm{MB-APEX}^2$$\sim($24$'')$$^2$/(18$''$)$^2$$\sim$1.8,
where $\Omega_\mathrm{MB-SEST}$ is the main-beam solid angle for SEST,
and $\Omega_\mathrm{MB-APEX}$, for APEX.  CO freeze-out on dust grains
may also affect the $N_{\mathrm{1.2mm}}$/$N_{\mathrm{^{13}CO}}$ ratio
in regions with low temperatures ($<$15\,K) and high densities
\citep[$>$10$^{4}$cm$^{-3}$;][]{kramer1999}.

\makeFigureDeltaN

\subsection{Mass of the ring}

%
%

Following Eqs. (\ref{equationLTEMass}) and (\ref{equationDustMass}) in
Appendices I and II, it is possible to obtain two different
measurements of the total mass of the ring using the $^{13}$CO(3-2)
line and the 1.2\,mm continuum emission independently.  The total
integrated emission in the $^{13}$CO(3-2) line is
$\sim$1.8x10$^7$\,K\,km\,s$^{-1}$\,arcsec$^2$, implying that the
derived total mass is 6.9$\times$10$^3$\,M$_\odot$.  The total flux
density in the 1.2\,mm continuum emission is $\sim$120\,Jy, giving a
total mass of $\sim$4.0$\times$10$^3$\,M$_\odot$.  Considering the
errors of the assumed physical parameters (see
Sect. \ref{sectionColumnDensityOfTheRing}), the mass estimates agree
well.

\makeTableSummary

\subsection{Identification of the $^{13}$CO clumps}
\label{13CO(3-2)Clumps}

%
%

As Figs. \ref{figure13COComponentes}b and \ref{figure13COComponentes}c
show, the $^{13}$CO(3-2) line emission is clumpy.  Using
CLUMPFIND\footnote{http://www.ifa.hawaii.edu/users/jpw/clumpfind.shtml}
\citep{williams1994}, 104 clumps ($^{13}$CO clumps) were
identified. They contain 65\% of the total integrated emission,
$\sim$1.2$\times$10$^7$\,K\,km\,s$^{-1}$\,arcsec$^2$.  CLUMPFIND
creates contours over the data, searches for peaks of emission to
locate clumps, and follows them down to the lower intensity contour.
To find these clumps, the algorithm was applied with a lower intensity
contour of three rms and with a contouring interval equal to twice the
rms.  Fictitious clumps were filtered out by the condition that
intensity peaks have to be higher than five times the rms, $\sim$5\,K.

\subsection{Evolution of clumps}
\label{sectionAssociationWithMillimeterAndInfraredEmission}

%
%

The evolution of the clumps can be investigated by associating the
$^{13}$CO clumps with the sources detected in 1.2\,mm continuum and
MSX observations.

%
%
 
The clumps are classified into three types, A, B, and C.  Type A
clumps have a counterpart seen in infrared and millimeter wavelengths,
type B clumps only have a counterpart in the 1.2\,mm continuum, and
type C clumps are only seen in $^{13}$CO.

%
%

The criterion for correlating the sources was to identify the centers
closer than $\sim$27$''$, 1.5\,beam sizes of the $^{13}$CO line
observations.  Since the 8.3\,$\mu$m MSX band is sensitive to the PAH
emission and to the photospheric emission from stars
\citep[e.g.][]{chavarria2008}, clumps not detected in all MSX bands
are considered to have no counterpart at infrared wavelengths.

%
%
About 6\% of the $^{13}$CO clumps are type A, 27\% of the clumps are
type B, and the remaining clumps, 67\%, are type C.

%
%

Owing to the sensitivity limit of the MSX observations, the percentage
of type A clumps is a lower limit to the number of clumps that are
forming stars, and the percentages of types B and C clumps are an
upper limit to the number of clumps that are not forming stars.

%
%

As the $^{13}$CO(3-2) line is more sensitive to lower densities than
the 1.2\,mm continuum (see Sect. \ref{sectionColumnDensityOfTheRing}),
type C clumps have to correspond to structures with low densities that
can represent early states of the clumps.

%
%

Assuming that type A clumps are embedded clusters with typical ages of
2.5\,Myr (Lada \& Lada 2003) and that the timescale for each
evolutionary stage is proportional to the number of clumps, type B
clumps have timescales of $\lesssim$11\,Myr and type C clumps
timescales of $\lesssim$28\,Myr.  Thus, the typical lifetime for these
clumps probably is $\lesssim$40\,Myr, an upper limit comparable with
the GMC lifetimes of $\sim$20-40 Myr
\citep{blitz1993,kawamura2009,miura2012}.

\subsection{Physical properties of the $^{13}$CO clumps}

%
%

Using Eq. \ref{equationLTEMass}, the derived masses of the clumps are
between 2.3 and 7.5$\times$10$^2$\,M$_\odot$, with an average value of
44\,M$_\odot$.  The lowest detected clump mass is 0.1\,M$_\odot$,
assuming the smallest dimension of 0.111\,km\,s$^{-1}$ $\times$
20\,arcsec $\times$ 20\,arcsec and lowest antenna temperature of 5\,K
(5 sigma level).

%
%

Clump diameters, $D_\mathrm{c}$, are estimated from the deconvolved
FWHM size of their emissions using Eq. \ref{equationSize} in Appendix
III.  The clumps have diameters of between 0.3 and 1.6\,pc with an
average of 0.9\,pc.

%
%

Using the masses and diameters, the mean column density,
N$_\mathrm{c}$, and the mean density, n$_\mathrm{c}$, are calculated
using Eqs.  \ref{equationColumnDensity} and \ref{equationDensity} in
Appendix III, respectively.  Clumps have column densities between
3$\times$10$^{20}$ and 2$\times$10$^{22}$\,cm$^{-2}$ with an average
value of 3$\times$10$^{21}$\,cm$^{-2}$, and densities between 10$^2$
and 10$^4$\,cm$^{-3}$ with an average value of
2$\times$10$^3$\,cm$^{-3}$.

%
%

\begin{figure}
\ifshowImage
\resizebox{\hsize}{!}{\includegraphics{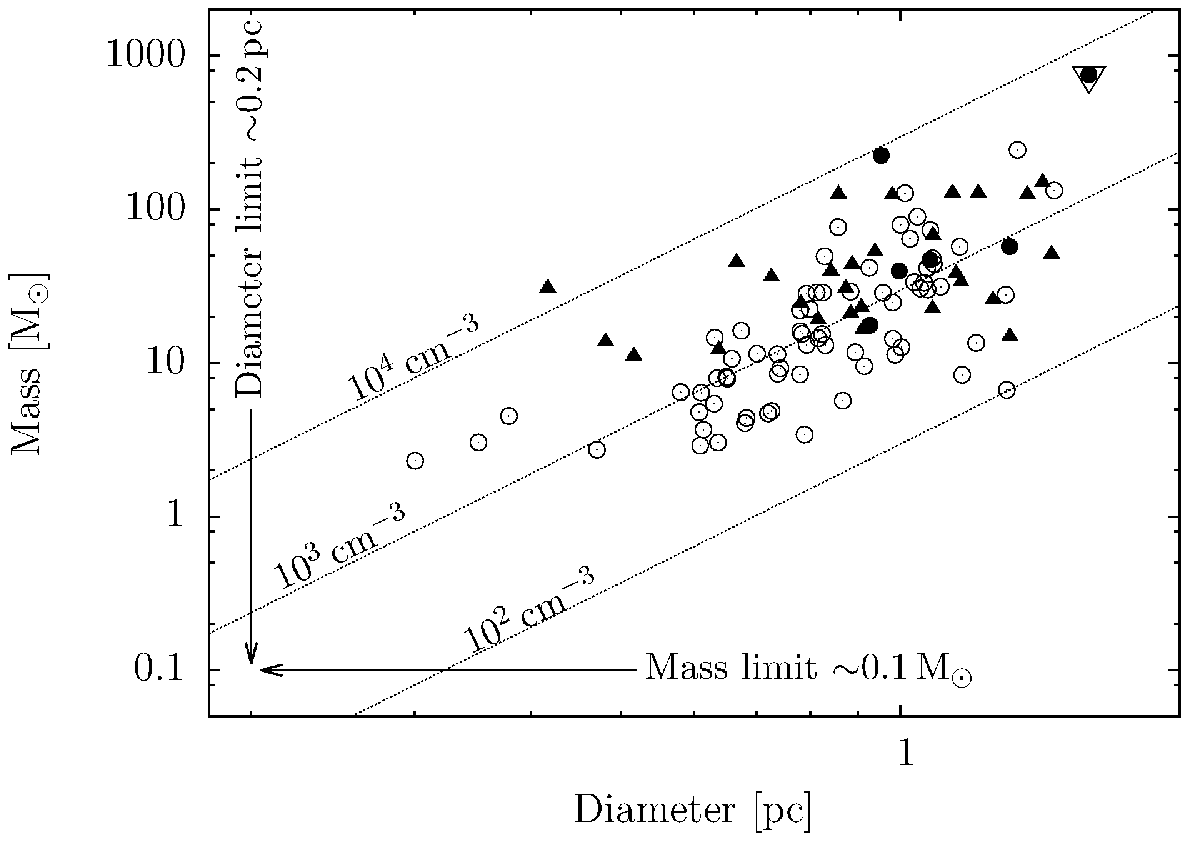}}
\fi
\caption{
Mass versus diameter for the clumps detected in the $^{13}$CO(3-2)
line.  The filled circles represent type A clumps, the filled
triangles show type B clumps, and the open circles plot the type C
clumps.  The open triangle indicates the clump associated with the
MSFR IRAS\,16562-3959.  The arrows show the minimal detectable diameter
and mass, $\sim$0.2\,pc and $\sim$0.1\,M$_\odot$.  The dotted lines
indicate mean densities at 10$^2$, 10$^3$, and 10$^4$\,cm$^{-3}$.
}
\label{figureMass-Diameter}
\end{figure}

%
%

Figure \ref{figureMass-Diameter} shows a plot of mass versus diameter
for the $^{13}$CO clumps, with lines to display different densities.
Type A clumps are the most massive and dense in average,
$\sim$1.9$\times$10$^2$\,M$_\odot$ and 3.2$\times$10$^3$\,cm$^{-3}$.
One example is the clump associated with the MSFR IRAS\,16562-3959, with
a mass of $\sim$7.5$\times$10$^2$\,M$_\odot$ and density of
6$\times$10$^3$\,cm$^{-3}$. In the other extreme, type C clumps are
the less massive and less dense on average, which is consistent with
the comments in
Sect. \ref{sectionAssociationWithMillimeterAndInfraredEmission}, where
we indicated that these clumps might represent an earlier state in the
clump formation.

%
%

\begin{figure}
\ifshowImage
\resizebox{\hsize}{!}{\includegraphics{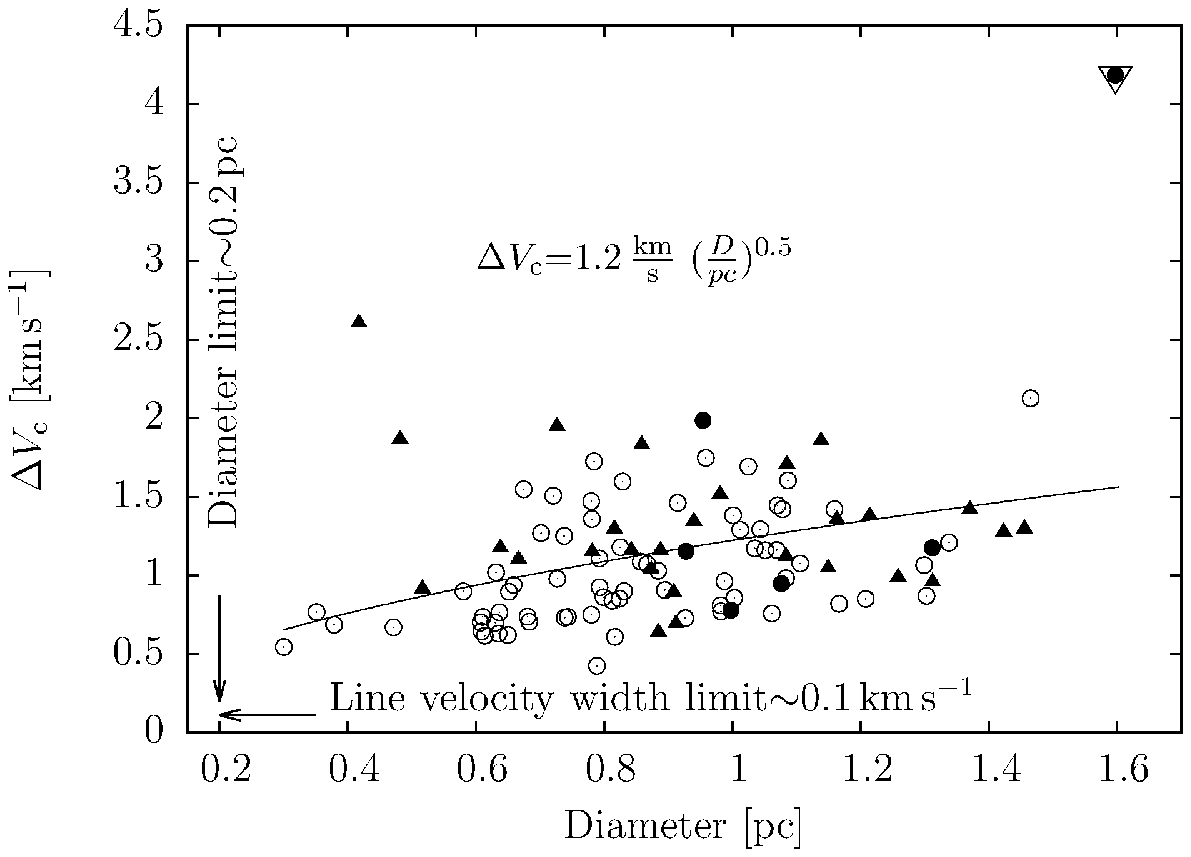}}
\fi
\caption{
Line velocity width versus diameter for the clumps detected in the
$^{13}$CO(3-2) line.  Arrows show observational detection limits for
line widths and diameters, $\sim$0.1~km\,s$^{-1}$ and
$\sim$0.2\,pc. The line shows the fit, $\Delta
V_\mathrm{c}$=1.2\,km\,s$^{-1}$ (D$_\mathrm{c}$\,pc$^{-1}$)$^{0.5}$.
The symbols are the same as in Fig. \ref{figureMass-Diameter}.
}
\label{figureLinewidth-Diameter}
\end{figure}

Figure \ref{figureLinewidth-Diameter} shows line velocity width
(FWHM), $\Delta V_\mathrm{c}$, versus diameter, $D_\mathrm{c}$, for
the clumps.  Line widths range between 0.4 and 4.2\,km\,s$^{-1}$ with
an average value of 1.2\,km\,s$^{-1}$. The relationship between line
width and diameter is given by
\begin{equation}
\qquad\qquad\qquad\qquad \Delta
V_\mathrm{c}\sim1.2\frac{\mathrm{km}}{\mathrm{s}}\left(\frac{D_\mathrm{c}}{\mathrm{pc}}\right)^{0.5}.
\end{equation}
This relationship is weak, with a correlation coefficient of 0.33
and dispersion of 0.5 km\,s$^{-1}$.  The clump associated with
IRAS\,16562-3959 shows the highest dispersion, which is explained by the
outflow identified in this clump (see Sect.
\ref{sectionUnresolvedMolecularOutflow}).

%
%

Table \ref{tableSummary} presents a summary of the physical properties
of the clumps, and Table \ref{table13COClumps} lists the individual
estimates for each clump.

%
%

The physical properties of identified clumps agree with observations
toward other GMCs.  For example, clumps within the GMC associated with
RCW 106 detected in the $^{13}$CO(1-0) line, C$^{18}$O(1-0) line, and
1.2\,mm continuum have diameters of 0.3 to 4\,pc, masses of 10 to
2$\times$10$^4$\,M$_\odot$, densities of 2$\times$10$^3$ to
6$\times$10$^4$\,cm$^{-3}$, and line velocity widths of 0.4 to
3\,km\,s$^{-1}$ \citep{mookerjea2004,bains2006,wong2008}.

%
%

\subsection{Gravitational stability}

%
%

To investigate whether clumps are gravitationally bound, their
physical properties were examined using two methods, assuming virial
equilibrium \citep[e.g.][]{bertoldi1992}, and assuming that they can
be modeled as Bonnor-Ebert spheres \citep{bonnor1956}.

%
%

Using the virial condition, Eq. \ref{equationAlpha} in Appendix IV
defines $\alpha_\mathrm{virial}$, which can be used as an indicator of
gravitational stability, because it is a measurement of the ratio of
the kinetic energy to the gravitational energy
\citep[e.g.][]{wong2008}.  For $\alpha_\mathrm{virial}$$>>$1, clumps
are not bound and must be confined by an external pressure to be in
hydrostatic equilibrium, for $\alpha$$\sim$1, clumps are in
equilibrium, and for $\alpha$$<<$1, clumps are unable to support
themselves against gravity.

%
%

\begin{figure}
\ifshowImage
\resizebox{\hsize}{!}{\includegraphics{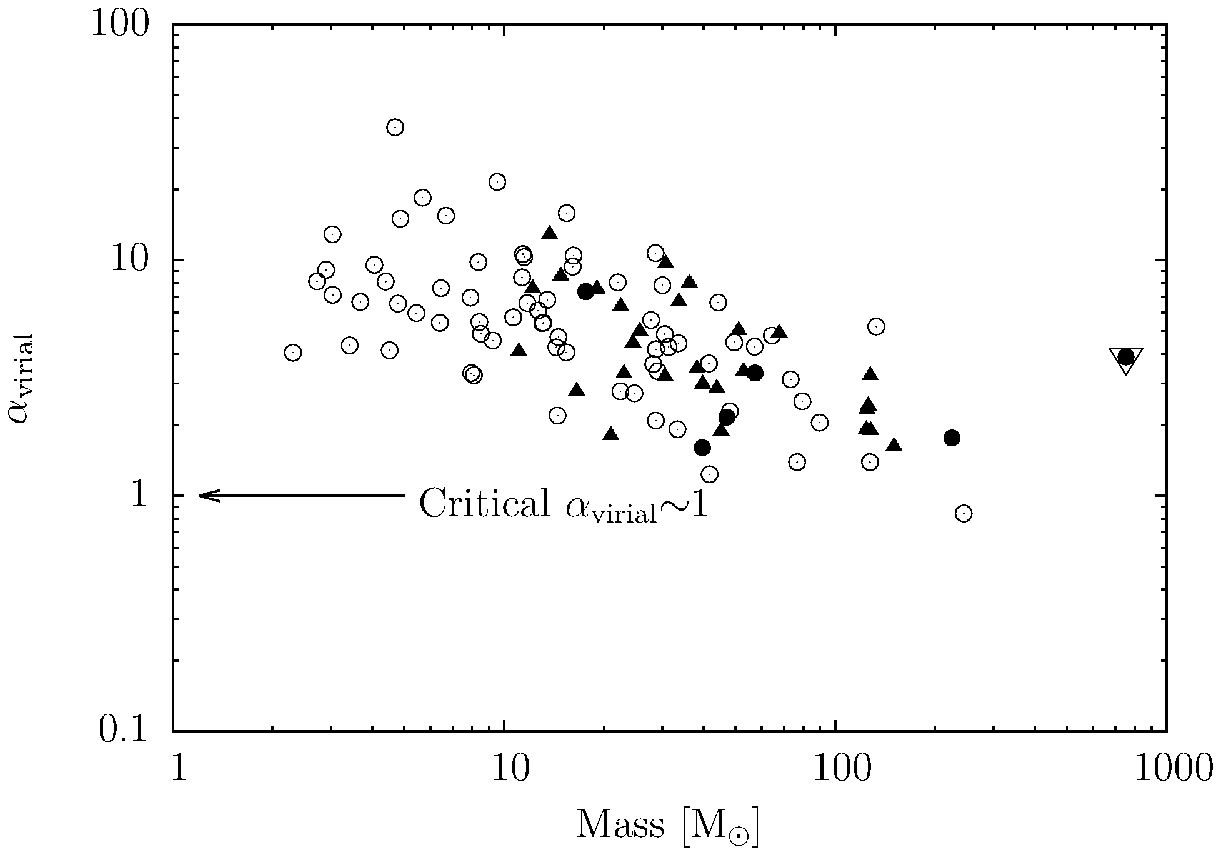}}
\fi
\caption{
$\alpha_\mathrm{virial}$ versus mass for the identified clumps. The
arrow indicates the critical values of $\alpha_\mathrm{virial}$,
$\sim$1.  The symbols are explained in Fig. \ref{figureMass-Diameter}.
}
\label{figureMassAlphaVirial}
\end{figure}

Figure \ref{figureMassAlphaVirial} displays $\alpha_\mathrm{virial}$
versus mass.  All clumps have $\alpha_\mathrm{virial}$$\gtrsim$1, thus
they must be in equilibrium or confined by an external pressure to be
in hydrostatic equilibrium.

%
%

%
%

\begin{figure}
\ifshowImage
\resizebox{\hsize}{!}{\includegraphics{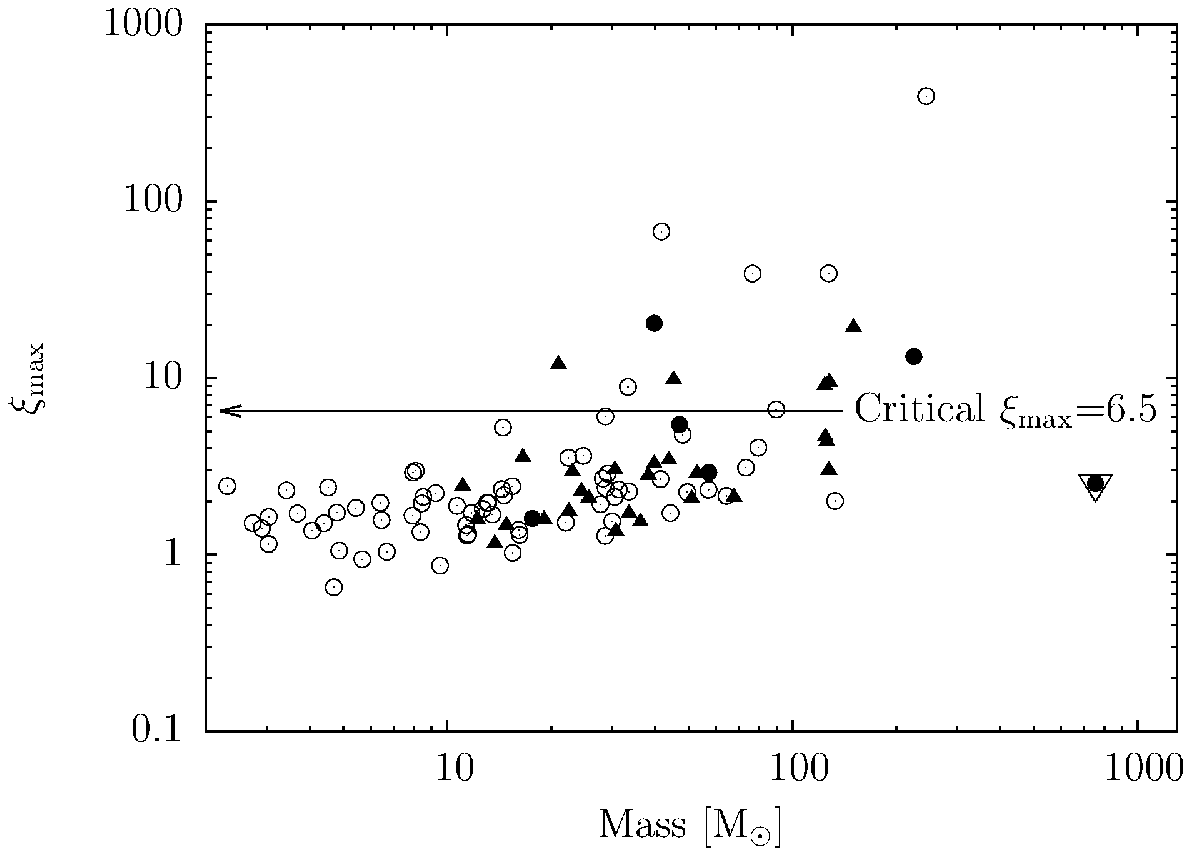}}
\fi
\caption{
Dimensionless radius $\xi_\mathrm{max}$ versus mass for the clumps
detected in the $^{13}$CO(3-2) line.  The arrow marks the critical
state for the Bonnor-Ebert sphere ($\xi_\mathrm{max}$=6.5).  The
symbols are explained in Fig. \ref{figureMass-Diameter}.
}
\label{figureXi-mass}
\end{figure}

\begin{figure}
\ifshowImage
\resizebox{\hsize}{!}{\includegraphics{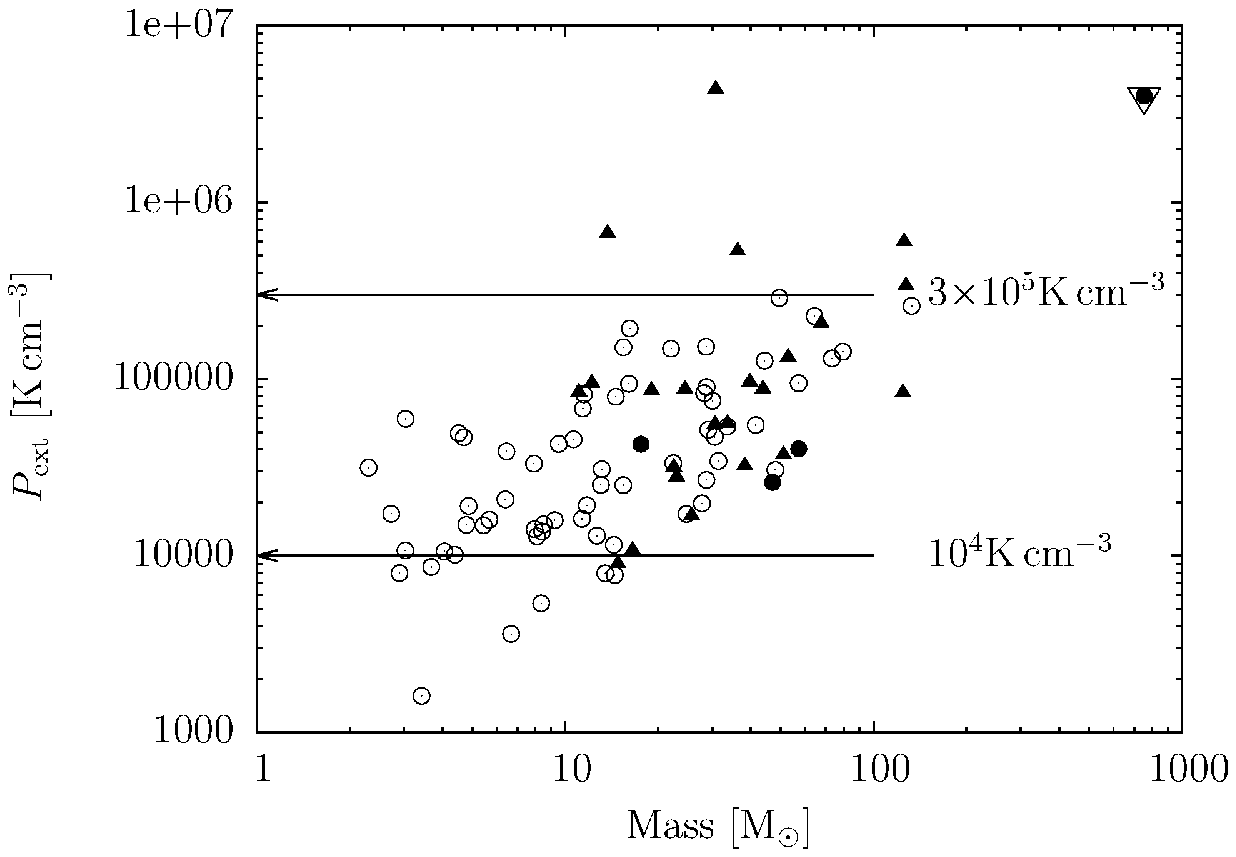}}
\fi
\caption{
External pressure $P_\mathrm{ext}$ versus mass for the clumps
identified in the $^{13}$CO(3-2) line with $\xi_\mathrm{max}$$<$6.5.
The arrows indicate the internal pressure estimated for the GMC
G345.5+1.5 using its physical properties
($\sim$3$\times$10$^5$\,K\,cm$^{-3}$), and the mean local interstellar
pressure ($\sim$10$^4$\,K\,cm$^{-3}$).  The symbols are explained in
Fig. \ref{figureMass-Diameter}.
}

\label{figurePext-mass}
\end{figure}

We employed a Bonnor-Ebert sphere analysis (see Appendix II) to
determine the dimensionless radius $\xi_\mathrm{max}$ and the external
pressures $P_\mathrm{ext}$.  When $\xi_\mathrm{max}$$>$6.5, the gas
sphere is susceptible to gravitational collapse. In the other case,
when $\xi_\mathrm{max}$$<$6.5, the gas sphere is thought to be stable
against collapse.  Figure \ref{figureXi-mass} displays the
dimensionless radius $\xi_\mathrm{max}$ versus mass for all clumps,
and Fig. \ref{figurePext-mass} shows the external pressure
$P_\mathrm{ext}$ versus mass for the stable clumps.

%
%

%
%

%
%

In general, $\xi$ seems to increase with mass, which means that with
increasing mass the likelihood increases that they are in collapsing
phase. However, most clumps are in hydrostatic equilibrium confined by
external pressure, which partly agrees with the virial equilibrium
analysis, since Bonnor-Ebert sphere analysis also shows clumps in a
stage susceptible to gravitational collapse.  These clumps are
indicated in Table \ref{table13COClumps}.

%
%

The clumps in a hydrostatic equilibrium require external pressures
of 1.6$\times$10$^3$ to 4$\times$10$^6$\,K\,cm$^{-3}$, with a median
value of $\sim$4$\times$10$^{4}$\,K\,cm$^{-3}$.  This pressure
represents an estimate of the pressure inside this GMC, which is
consistently higher than the the mean local interstellar pressure
\citep[$\sim$$10^4\,\mathrm{K\,cm^{-3}}$;][]{bloemen1987}, but lower
than the pressure required to prevent the GMC from collapsing
under its own weight in virial equilibrium
\citep[e.g.][]{bertoldi1992},
\begin{equation}
\ \ \ 
P_\mathrm{GMC}\sim-\frac{W_\mathrm{GMC}}{3V_\mathrm{GMC}}=\frac{1}{5}\frac{G\,M^2_\mathrm{GMC}}{V_\mathrm{GMC}R_\mathrm{GMC}}\sim3\times10^5\,\mathrm{K\,cm^{-3}},
\end{equation}
where $V_\mathrm{GMC}$, $M_\mathrm{GMC}$ and $R_\mathrm{GMC}$ are the
volume ($\sim$$4/3\pi R^{3}_\mathrm{GMC}$), mass
($\sim$6.5$\times$10$^5$\,M$_\odot$), and radius ($\sim$34\,pc) of the
GMC \citep{lopez2011}.  This estimation for
$P_\mathrm{GMC}$ is similar to the typical internal pressures of GMCs,
$\sim$10$^5$ K\,cm$^{-3}$ \citep[e.g.][]{blitz1993}.

%
%

\begin{figure*}
\ifshowImage
\resizebox{\hsize}{!}{\includegraphics{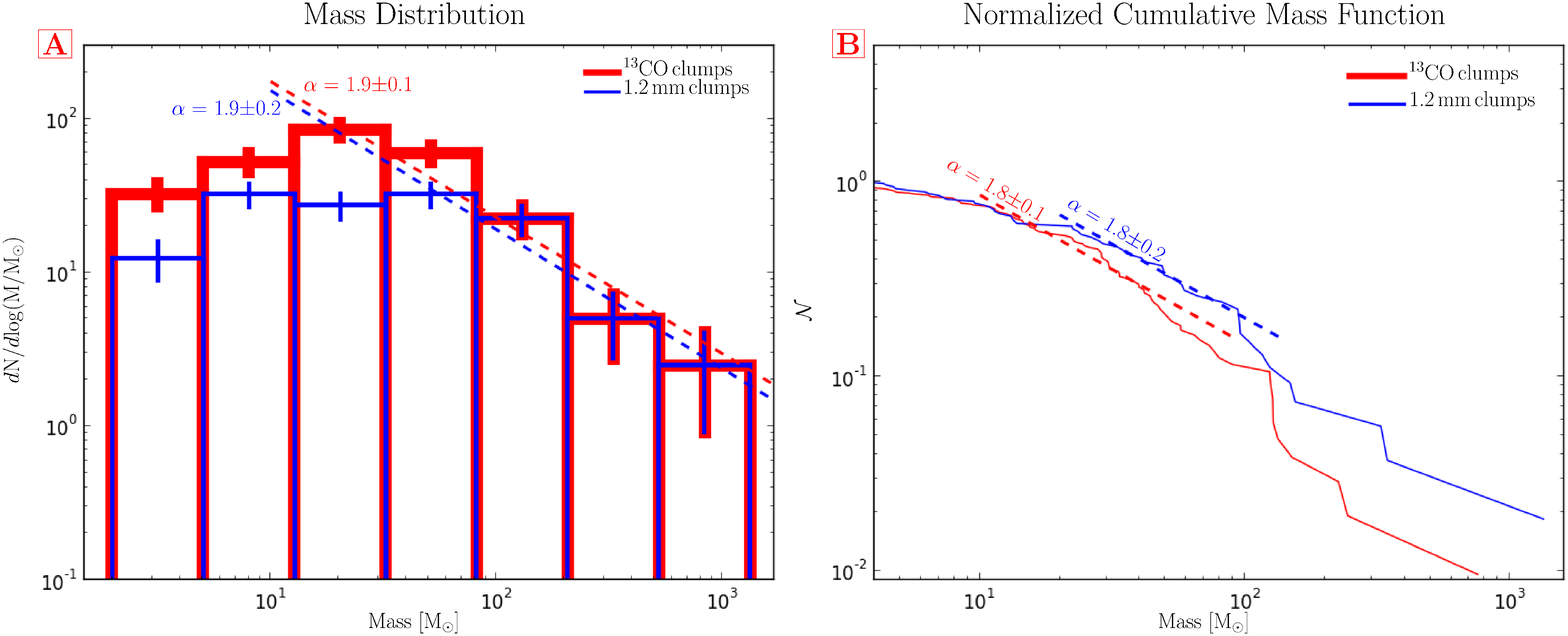}}
\fi
\caption{
Panel a: histogram of the mass distributions for the $^{13}$CO clumps
(solid red lines) and 1.2\,mm clumps (solid blue lines).  The dashed
lines indicate the fits with $\alpha$=1.9$\pm$0.1 for $^{13}$CO clumps
and $\alpha$=1.9$\pm$0.2 for 1.2\,mm clumps.  Panel b: the normalized
cumulative mass functions for the $^{13}$CO clumps (solid red lines)
and 1.2\,mm clumps (solid blue lines).  The dashed lines indicate the
fits with $\alpha$=1.8$\pm$0.1 for $^{13}$CO clumps and
$\alpha$=1.8$\pm$0.2 for 1.2\,mm clumps.
}
\label{figureCMD}
\end{figure*}

\subsection{Comparison with 1.2\,mm clumps}

%
%

\begin{table}

\caption{
Summary of the physical properties of the 54 1.2\,mm clumps identified
in the ring G345.45+1.50 \citep{lopez2011}.
}

\label{tableSummary1.2mm}
\begin{center}
\begin{tabular}{lccc}
\hline\hline\\
Parameter            & Range      & Average  & Median \\
\hline\\
Diameter [pc]           & 0.2-0.6&    0.3 & 0.3\\
 Mass   [M$_\odot$]         & 3.8-1.3$\times$10$^3$ & 75 & 27\\
Density [cm$^{-3}$]        & 8$\times$10$^3$-3$\times$10$^5$&  8$\times$10$^4$& 5$\times$10$^4$\\
\hline
\end{tabular}
\end{center}

\end{table}

%
%

A summary of the physical properties of the 54 1.2\,mm clumps
identified in the ring is shown in Table
\ref{tableSummary1.2mm}. Compared with Table \ref{tableSummary}, the
typical masses of the 1.2\,mm clumps and the $^{13}$CO clumps are
similar, but the 1.2\,mm clumps are on average about three times
smaller than the $^{13}$CO clumps.  This is consistent with the
greater extension of the $^{13}$CO(3-2) line emission.

%
%

The mass distributions for the 1.2\,mm and $^{13}$CO clumps are shown
in Fig. \ref{figureCMD}a.  From this plot, the completeness limits of
the mass distributions are estimated to be $\sim$33\,M$_\odot$ and
$\sim$13\,M$_\odot$ for 1.2\,mm and $^{13}$CO clumps, respectively.
Bin errors are calculated as $\sqrt{\Delta N/\Delta\log(M/M_\odot)}$,
where $\Delta N$ is the number of clumps in the constant logarithmic
mass interval $\Delta \log(M/M_\odot)$, $\sim$0.4.  When we consider
clumps with masses higher than the completeness limits, the fits of
the power law function
\begin{equation}
\ \ \ \ \ \ \ \ \ \ \ \ \ \ \ \ \ \ \ \ \ \ \ \ \ \ \ \ \ 
dN/d \log(M)\propto M^{-\alpha+1}
\end{equation}
in the two distributions are consistent with a value for
$\alpha$ of $\sim$1.9.

%
%

As the sample of 1.2\,mm clumps is smaller than 70, its mass
distribution can be affected by the binning \citep[e.g.][]{munoz2007}.
To check the value of $\alpha$ equal to 1.9 estimated from the mass
distributions, the normalized cumulative mass functions of the two
clump samples were calculated and are shown in Fig. \ref{figureCMD}b.
To avoid the invariable increase of the slope at the end of the
normalized cumulative mass function, as  explained by
\cite{munoz2007}, 20\% of the total cloud mass was excluded from the
fittings.  The clumps with masses lower than the completeness limit
were also excluded.  In this range of masses, the relationship between
$\alpha$ and the cumulative mass function is approximated as
\citep{munoz2007}
\begin{equation}
\ \ \ \ \ \ \ \ \ \ \ \ \ \ \ \ \ \ \ \ \ \ \ \ \ \ \ \ \ 
\mathcal{N}(M) \propto M^{-\alpha+1}.
\end{equation}

%
%

The normalized cumulative mass function and the mass distribution both
show consistent values for $\alpha$ equal to $\sim$1.9.  This result
also agrees with previous studies of clumps in high-mass star-forming
regions made at spatial resolutions between 0.1 and 0.8\,pc
\citep[e.g.][]{munoz2007,lopez2011,mookerjea2004,
  kerton2001,tothill2002,kramer1998,wong2008}. Using CO isotope lines
and continuum emission, these studies determined clump mass
distributions with $\alpha$ of between 1.6 and 1.9.  From numerical
models of self-gravity molecular cloud dynamics including different
levels of turbulent support, the observed range of $\alpha$ is
explained by different grades in the domination of the gravity; the
clump mass distribution becomes shallower when gravity increases
\citep{klessen2001}. With these values for $\alpha$, additional
fragmenting processes are needed to obtain the core mass function
\citep[e.g.][]{nutter2007} and the star initial mass function
\citep[IMF; e.g.][]{kroupa2002,kroupa2007}, which have an $\alpha$
closed to 2.35. These additional fragmenting processes are consistent
with the observations, where the vast majority of stars form in
binaries or higher-order multiple system \citep{goodwin2007}.

%
%

Studies at smaller scales ($<$0.1\,pc) and lower masses
($<$10\,M$_\odot$) show that the core mass function has similar slopes
to the stellar IMF \citep[e.g][]{motte1998,alves2007,nutter2007},
arguing that the fragmentation process at this scale sets the IMF.  In
contrast to this hypothesis, \cite{goodwin2008} proved that the model
in which all stars and brown dwarfs form in multiple system from a
core mass distribution provides a very good fit to the IMF.  Our
observations have a spatial resolution of 0.2\,pc, therefore we cannot
determine the mass function on scales $<$0.1\,pc.

%
%

An exception to the form of the clump mass function was found by
\cite{reid2006}. They rederived the clump mass function in some GMCs
and found a double power-law with a slope at the high-mass end of 2.4,
but we fail to see any indication for such a distribution.

\subsection{Kinematic structure}
\label{sectionExpandingRing}

%
%

\begin{figure}
\ifshowImage
\resizebox{250pt}{!}{\includegraphics{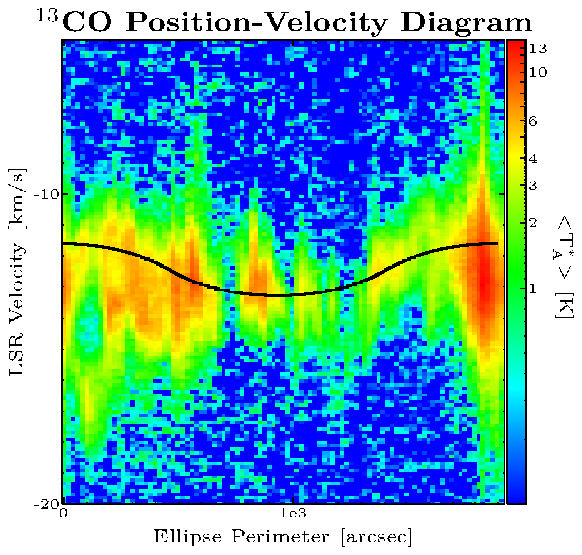}}
\fi
\caption{
Position-velocity diagram of the $^{13}$CO(3-2) line along the
elliptical curve of the expanding ring model in the X-Y plane (see
Fig. \ref{figure13COComponentes}a); the black line shows the
position-velocity diagram for the model in
Sect. \ref{sectionExpandingRing}.
}
\label{figureExpandingRing}
\end{figure}

The $^{13}$CO(3-2) line emission of the whole region shows the
characteristics of an expanding ring-like structure, as shown in
Figs. \ref{figure13COComponentes}a and \ref{figureExpandingRing},
where we display the velocity-integrated emission and
position-velocity diagram, respectively.  To characterize this
structure, we used a new method: a model assuming a ring expanding at
constant velocity rotated at an angle $\alpha$ in the plane X-Z, and
rotated at an angle $\beta$ in the observed spatial plane (X-Y plane).
Thus
\begin{equation}
X(\theta)=X_0+R(\cos\theta\cos\alpha\cos\beta-\sin\theta\sin\beta)
\end{equation}
\begin{equation}
Y(\theta)=Y_0+R(\cos\theta\cos\alpha\sin\beta+\sin\theta\cos\beta)
\end{equation}
\begin{equation}
V_\mathrm{z}(\theta)=V_0+V\cos\theta\sin\alpha,
\end{equation}
where $\theta$ is the angular position in the ring, between 0 and
2$\pi$, $X(\theta)$ and $Y(\theta)$ are the projected spatial
position, $V_\mathrm{z}(\theta)$ is the velocity in the Z direction,
$X_0$ and $Y_0$ define the spatial center, $V_0$ is the velocity
center, $R$ is the radius, and $V$ is the expansion velocity.

%
%

%
%

The fit yields a center at RA=16:59:26, Dec=$-$40:04:30 (J2000) and an
LSR velocity= $-$12.4\,km\,s$^{-1}$, with a radius and expansion
velocity of 3.4\,pc and 1.0\,km\,s$^{-1}$, respectively.  The
expansion timescale is $\sim$3$\times$10$^6$\,yr ($t$=$R$/$V$), and
considering a total mass of 6.9$\times$10$^3$\,M$_\odot$, the total
kinematic energy is $\sim$7$\times$10$^{46}$\,erg
($E$=$M$\,$V$$^2$/2).  Table \ref{tableRing} summarizes these
parameters, and Figs. \ref{figure13COComponentes}a and
\ref{figureExpandingRing} show the expanding ring model over the
integrated velocity image and over the position-velocity diagram of
the $^{13}$CO line, respectively.

%
%

\begin{table}
\centering
\caption{Characteristics of the expanding ring.}
\label{tableRing}
\begin{tabular}{l l}
\hline\hline
Parameter &  Value \\
\hline\\
Spatial center   &    16:59:26 $-$40:04:30\,(J2000)\\
LSR velocity center  &  $-$12.4\,km\,s$^{-1}$   \\
Radius           & 3.4\,pc \\
Expansion velocity & 1.0\,km\,s$^{-1}$\\
$\alpha$ & 57\,deg\\
$\beta$  & $-$175\,deg \\
Mass           &  6.9$\times$10$^{3}$\,M$_\odot$\\
Energy         &  7$\times$10$^{46}$\,erg\\
Expansion time & 3$\times$10$^6$\,yr\\
\hline
\end{tabular}
\end{table}

%
%

Figure \ref{figureMGPS-MSX-SPITZER}d shows the 35.6\,cm continuum
emission overlaid with the elliptical curve of the ring model in the
X-Y plane.  The 35.6\,cm source J165920-400424 is located within the
expanding ring, at $\sim$30~arcsec from its center, and without an
infrared counterpart, as was discussed in
Sect. \ref{sectionStarFormationProcessAlongTheRing}.  Thus it is
possible that the expansion of the ring is caused by a supernova
explosion, where J165920-400424 can be a pulsar
\citep[e.g.][]{whiteoak1992}, resulting from the gravitational
collapse of a massive star.  This explanation needs to be tested in
future observations (see
Sect. \ref{sectionStarFormationProcessAlongTheRing}).

%
%

\begin{table}
\centering
\caption{Characteristics of the outflow IRAS\,16562-3959.}
\label{tableOutflows}
\begin{tabular}{l l }
\hline\hline
Parameter &  Value\\
\hline\\
Spatial center  (J2000)    &16:59:41 \\
                          &$-$40:03:37 \\
LSR velocity center [km\,s$^{-1}$]   & $-$11.3 \\
Projected size [pc]           & 0.4$^\dagger$\\
Projected velocity range [km~s$^{-1}$]   & 38.4\\
Mass [M$_\odot$]           & 13\\
Inclination [deg]& 41$^\dagger$\\
Energy [erg] &   7$\times$10$^{45}$\\
Expansion time [yr] & 10$^4$$^\ddagger$\\
Mass outflow rate [M$_\odot$\,yr$^{-1}$] &  2$\times$10$^{-3}$$^\ddagger$\\
Luminosity  [L$_\odot$]         & 5-6$\times$10$^4$ $^\dagger$$^\star$\\
\hline
\end{tabular}
\begin{list}{}{}
\item[$^\dagger$]From \cite{guzman2010}
\item[$^\ddagger$]Using the size estimated by  \cite{guzman2010}
\item[$^\star$]From \cite{lopez2011}
\end{list}{}{}
\end{table}

%
%

\subsection{Molecular outflow in IRAS\,16562-3959}
\label{sectionUnresolvedMolecularOutflow}

The outflow associated with the infrared source IRAS\,16562-3959
(hereafter outflow IRAS\,16562-3959) was first identified as a
collimated jet by \cite{guzman2010} using 3.5-21~cm continuum
emission.  They determined an outflow size of $\sim$0.4~pc (47$''$ at
1.8~kpc).

In the $^{12}$CO and $^{13}$CO lines observations shown in
Fig. \ref{figure13COComponentes}e, this molecular outflow is
identified by the broad line width and clear line wings.

From the $^{12}$CO(3-2) line, the velocity range of the outflow is
between $-$31.1 and 7.3~km~s$^{-1}$ with emission $>$3$\sigma$, and
centered at $-$11.3~km~s$^{-1}$ calculated by fitting a Gaussian shape.
Using the $^{13}$CO(3-2) line and eliminating the ambient cloud
component by fitting a Gaussian shape to the integrated spectrum of
the central area of the clump within a radius of 23.5$''$, the mass
and projected energy of the outflow are 13~M$_\odot$ and
7$\times$10$^{45}$~erg, respectively.

Using the size determined from centimeter continuum emission by
\cite{guzman2010}, it is possible to calculate an expansion time of
10$^4$~yr and a mass outflow rate of 10$^{-3}$~M$_\odot$~yr$^{-1}$.
Using the correlation luminosity-mass outflow rate estimated by
\cite{shepherd1996}, the driving source IRAS\,16562-3959 has to have a
luminosity of $\sim$10$^3$-10$^4$\,$L_\odot$, which is comparable with
the estimate obtained by integrating the spectral energy distribution
(SED), $\sim$5-6$\times$10$^4$~L$_\odot$ \citep{guzman2010,lopez2011}.

All characteristics of this outflow are summarized in Table
\ref{tableOutflows}.

%
%

\subsection{Filamentary structures}
\label{sectionFilamentaryStructures}

At the edge of the ring, five filamentary structures are discovered,
as indicated over the $^{13}$CO line image shown in
Fig. \ref{figure13COComponentes}a.  These structures are modeled as
\begin{equation}
X(r)=X_0 + r\,\cos(\theta)
\end{equation}
\begin{equation}
Y(r)=Y_0 + r\,\sin(\theta)
\end{equation}
\begin{equation}
V_\mathrm{z}(r)= V_0 + r\,\nabla V,
\end{equation}

where $r$ is the position in the filament, between $-S/2$ and $S/2$
with $S$ as the spatial size, $X(r)$ and $Y(r)$ are the projected
spatial position, $\theta$ is the direction angle in the observed
spatial plane (X and Y axes), $V_\mathrm{z}(r)$ is the velocity in the
Z direction, $\nabla V$ is the projected velocity gradient, $X_0$ and
$Y_0$ define the spatial center, and V$_0$ is the velocity center.

Using $^{13}$CO emission, the sizes of these structures are of between
2.2 and 5.1\,pc, their widths of between 0.35 and 0.93\,pc, and their
masses of between $\sim$100 and 250\,M$_\odot$.  Considering their
masses, sizes, and spatial widths, their column density averages are
between 3.8$\times$10$^{21}$ and 7.3$\times$10$^{21}$\,cm$^{-2}$.
Almost all filaments show small velocity gradients
($<$0.5\,km\,s$^{-1}$\,pc$^{-1}$), except for filament 5, which has a
velocity gradient of 2.5\,km\,s$^{-1}$\,pc$^{-1}$. Table
\ref{tableFilaments} summarizes the characteristics estimated for
these filaments, and in Fig. \ref{figureFilaments} we show their
$^{13}$CO line position-velocity diagrams.

Since these filaments are not associated with infrared sources and
almost all of them have small velocity gradients, they probably
originate in the scenario where the low-density material in the
molecular clouds first collapses into filamentary structures, which
later fragment into clumps and cores, where the density reaches the
critical level to form stars.  This scenario has been outlined by
observations toward regions of star formation
\citep[e.g.][]{molinari2010}.  For example, \cite{goldsmith2008} found
filaments of 0.8-1.0\,pc lenght with column densities of
$\sim$10$^{21}$~cm$^{-2}$ in Taurus, which are similar to physical
properties determined for the identified filaments here.

Using Herschel observations, several authors have determined a typical
width of $\sim$0.1\ pc for the filaments in molecular clouds
\citep[e.g.][]{alvesDeOliveira2014, arzoumanian2011,polychroni2013}.
Here, the widths of the filaments have an average of $\sim$0.6\ pc,
similar to other studies \citep[e.g.][]{hennemann2012,contreras2013},
but this discrepancy might be caused by the lower spatial resolution
in our observations.

%
%

\begin{figure*}
\ifshowImage
\resizebox{\hsize}{!}{\includegraphics{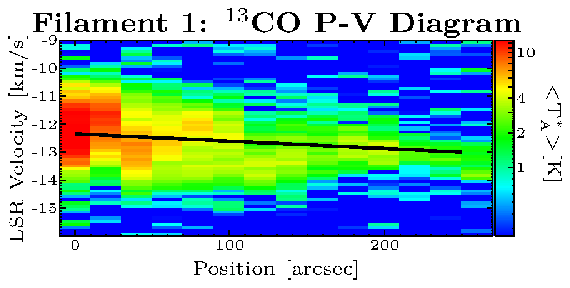}\includegraphics{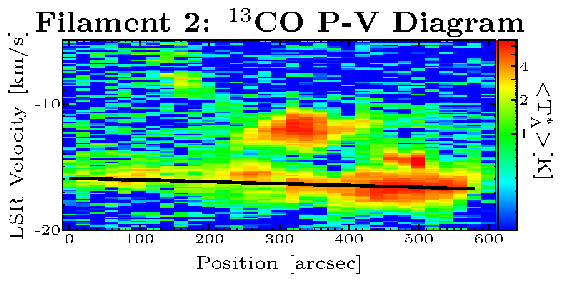}}
\resizebox{\hsize}{!}{\includegraphics{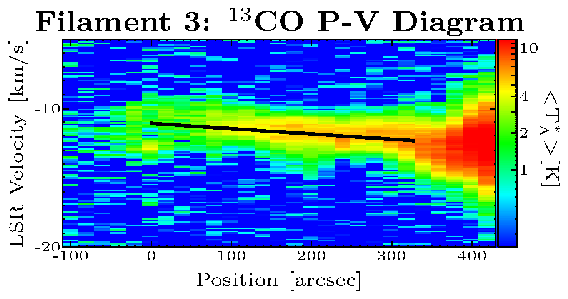}\includegraphics{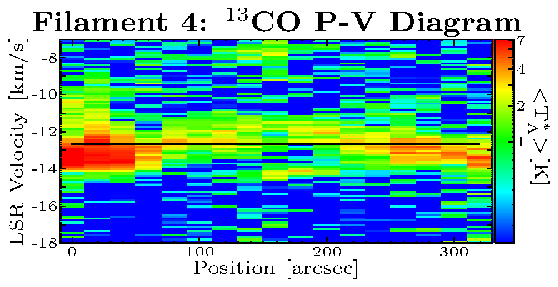}}
\centering
\resizebox{250pt}{!}{\includegraphics{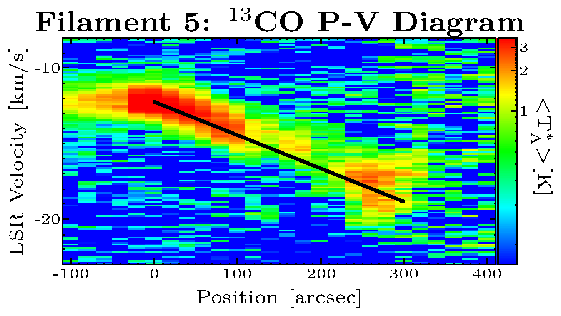}}
\fi
\caption{
$^{13}$CO(3-2) line position-velocity diagrams for the five filaments.
  The black lines indicate the position-velocity diagrams of the
  filament models, see Sect. \ref{sectionFilamentaryStructures}.
\label{figureFilaments}
}
\label{figureFilaments}
\end{figure*}

%
%

\begin{table*}
\centering
\caption{Characteristics of the filaments.}
\label{tableFilaments}
\begin{tabular}{l l l l l l c}
\hline\hline
Parameter & Filament 1 & Filament 2 & Filament 3 & Filament 4 & Filament 5  & Average \\
\hline\\
Spatial Center (J2000)                               & 16:58:56         & 16:59:49           & 16:59:54 &16:58:55 &   16:59:13.79  \\
                                                     & $-$40:05:36       & $-$40:06:00       & $-$39:59:34& $-$40:09:36&  $-$39:57:27 & \\
LSR velocity center [km\,s$^{-1}$]                    & $-$12.7           & $-$16.3           & $-$11.7    & $-$12.6&     $-$15.5     & \\
Size [pc]                                            &  2.2              & 5.1               & 2.9      &2.8 &                2.6  &  3.1\\
Projected velocity gradient [km\,s$^{-1}$\,pc$^{-1}$ ] &  0.3              & 0.2               & 0.5      &0.0&                2.5   & 0.7\\
Spatial width (FWHM) [pc]                        &  0.35               & 0.67               &0.65         &0.51  &        0.93    & 0.62\\
Velocity width (FWHM) [km\,s$^{-1}$]                  &   2.5             & 2.1              & 2.6       &2.2 &        2.5     & 2.4\\
Mass [10$^2$\,M$_\odot$]                                      &  1          & 2.4 &  2.2& 1.9& 2.5 &  2.0\\
Column density average [10$^{21}$\,cm$^{-2}$]                        &7.1             & 3.8    & 6.4  & 7.3 &  5.6 & 6.0 \\

\hline
\end{tabular}
\end{table*}

%
%

\section{Conclusions}
\label{sectionConclusions}

The ring G345.45+1.50 was observed in the $^{13}$CO(3-2) line. We list
our conclusions below.

\begin{itemize}

\item The ring contains a total mass of 6.9$\times$10$^3$\,M$_\odot$,
  which agrees well with results from previous observations in the
  1.2\,mm continuum emission.

\item The ratio of the column density estimated from the 1.2\,mm
  continuum to that estimated from the $^{13}$CO(3-2) line varies
  between 0.1 and 10, with an average value of $\sim$1.0. The column
  density estimated from the 1.2\,mm continuum is higher than that
  estimated from the $^{13}$CO(3-2) line toward sites with high
  density.  It is possible that the 1.2\,mm continuum emission is more
  optically thin than the $^{13}$CO(3-2) line toward dense regions.

\item The ring is expanding with a velocity of 1.0\,km\,s$^{-1}$ and
  has an expansion timescale of 3$\times$10$^6$\,yr and a total energy
  of 7$\times$10$^{46}$\,erg.  This expansion might have been produced
  by a supernova explosion. This hypothesis is supported by the
  presence of a 35.6\,cm source, J165920-400424, in the spatial center
  of the ring.  This source does not have an infrared counterpart and
  might be a pulsar, that remained from the gravitational collapse of
  a massive star.  This needs to be tested in future observations.

\item From the $^{13}$CO(3-2) line, the ring is composed of 104 clumps
  with diameters of between 0.3 and 1.6\,pc, masses of between 2.3 and
  7.5$\times$10$^2$\,M$_\odot$, and densities of between 10$^{2}$ and
  10$^{4}$\,cm$^{-3}$. About 6\% of them show clear signs of star
  formation, which allowed us to estimate a typical lifetime for these
  clumps of $\lesssim$40\,Myr.

\item Assuming that clumps can be modeled as Bonnor-Ebert spheres,
  $\sim$13\% of the clumps are gravitationally unstable, while most of
  the clumps require an external pressure with a median value of
  4$\times$10$^4$\,K\,cm$^{-3}$ to be in hydrostatic equilibrium.

\item In the region, the outflow associated with the MSFR
  IRAS\,16562-3959 was identified, with a velocity range of
  38.4~km~s$^{-1}$, a mass of 13~M$_\odot$, and a kinetic energy of
  7$\times$10$^{45}$~erg.

\item At the edge of the ring, five filamentary structures are found
  with lengths of between 2.2 and 5.1\,pc and with masses of between
  10$^2$ and 2.5$\times$10$^2$~M$_\odot$.

\end{itemize}

\begin{acknowledgements}

C.L.  acknowledges partial support from the GEMINI-CONICYT FUND,
project number 32070020, and ESO-University of Chile Student
Fellowship.  This work was supported by the Chilean Center for
Astrophysics FONDAP N$^\circ$ 15010003 and by Center of Excellence in
Astrophysics and Associated Technologies PFB 06.  LB acknowledges
support from CONICYT Project PFB06.  IdG acknowledges the Spanish
MINECO grant AYA2011-30228-C03-01 (co-funded with FEDER fund).

\end{acknowledgements}

\bibliographystyle{aa}

\bibliography{bibliography}

\onecolumn

\clearpage

%
%

\section{Appendix I: LTE column density and mass estimated  from the $^{13}$CO(3-2) line}

Considering the rotational transition of a linear molecule from the
upper level $\mathrm{u}$ to the lower level $\mathrm{l}$, the column
density of the lower level is \citep{wilson2009}
\begin{equation}
N_\mathrm{l}=93.5\,\mathrm{cm^{-2}}\,\frac{g_\mathrm{l}(\nu/\mathrm{GHz})^3}{g_\mathrm{u}
  (A_\mathrm{u\,l}/\mathrm{s^{-1}})}\frac{1}{1-\exp(-h\nu/k\,T_\mathrm{K})}\int\tau
\mathrm{(dv/km\,s^{-1})},
\end{equation}
where $T_\mathrm{ex}$ is the excitation temperature, which is
approximated to the kinetic temperature $T_\mathrm{K}$, $\nu$ is the
frequency of transition, $\tau$ is the optical depth,
$A_\mathrm{u\,l}$ is the Einstein A coefficient of the radiative
transition between the upper and lower levels, $\mathrm{v}$ is the
Doppler velocity, $g_\mathrm{l}$ and g$_\mathrm{u}$ are the
statistical weights of the states.  Assuming that $\tau$$<<$1 and the
source observed fills the main beam, we approximate
$T_\mathrm{K}\,\tau \sim T_\mathrm{MB}\sim T_\mathrm{b}$, where
$T_\mathrm{MB}$ is the main-beam brightness temperature and
$T_\mathrm{b}$ is the brightness temperature. Thus
\begin{equation}
N_\mathrm{l}=93.5\,\mathrm{cm^{-2}}\,\frac{g_\mathrm{l}(\nu/\mathrm{GHz})^3}{g_\mathrm{u}
  (A_\mathrm{u\,l}/\mathrm{s^{-1}})}\frac{1}{1-\exp(-h\,\nu/k\,T_\mathrm{K})}\frac{1}{T_\mathrm{K}}\int
T_\mathrm{b} \mathrm{(dv/km\,s^{-1})}.
\end{equation}

Assuming that all energy levels are populated under LTE conditions,
the fraction of the total population in a particular state, J, is
given by
\begin{equation}
N(J)/N(total) \sim (2J+1) \exp\left(- \frac{h B_\mathrm{e} J(J+1)}{k
  T_\mathrm{K}}\right) \frac{h\,B_\mathrm{e}}{k\,T_\mathrm{K}},
\end{equation}  
for $h\,B_\mathrm{e}$$<<$$k\,T_\mathrm{K}$, where $B_\mathrm{e}$ is
the rotational constant.  Thus, the total density is given by
\begin{equation}
N(total)=93.5\,\mathrm{cm^{-2}}\,\frac{g_\mathrm{l}}{g_\mathrm{u}}
\frac{(\nu/\mathrm{GHz})^3}{(A_\mathrm{u\,l}/s^{-1})} \frac{k}{h
  B_\mathrm{e}}\frac{1}{2J+1} \exp\left(\frac{h
  B_\mathrm{e}}{k}\frac{J(J+1)}{T_\mathrm{K}}\right)
\frac{1}{1-\exp(-h\,\nu/k\,T_\mathrm{K})} \int T_\mathrm{B}
\mathrm{(dv/ km\,s^{-1})}.
\end{equation}

For the emission transition J+1 to J, $A_\mathrm{u\,l}$ is given by
\begin{equation}
A_\mathrm{u\,l}= A_\mathrm{J}
=1.165\times10^{-11}\,\mathrm{s^{-1}}(\mu_D/\mathrm{debye})^2
(\nu/\mathrm{GHz})^3 \frac{J+1}{2J+3},
\end{equation}

$g_\mathrm{l}=2J+1$ and $g_\mathrm{u}=2J+3$. Thus 

\begin{equation}
N(total)=80.3\times10^{11}\,\mathrm{cm^{-2}}\,\frac{1}{(\mu_D/\mathrm{debye})^2}
\frac{k}{h B_\mathrm{e}}\frac{1}{J+1} 
\exp\left(\frac{h B_\mathrm{e}}{k}\frac{J(J+1)}{T_\mathrm{K}}\right) 
 \frac{1}{1-\exp(-h\,\nu/k\,T_\mathrm{K})}
\int T_\mathrm{B} \mathrm{(dv/ km\,s^{-1})}.$$
\end{equation}

For $^{13}$CO, the $B_\mathrm{e}$ is 5.5101$\times$10$^4$\,MHz, and
$\mu_D$, 0.11046\,debye\footnote{http://www.splatalogue.net/}.
Thus the total column density from the $^{13}$CO(3-2) transition
(330.587960\,GHz) is

\begin{equation}
\label{equationLTE}
\ \ \ \ \ \ \ \ \ \ \ \ \ \ \ \ \ \ \ \ \ \ \ \ \ \ \ \ \ \ \ \ \ \ \ \ \ \ \ \ \ \ \ N(total)=8.30\times10^{13}\,\mathrm{cm^{-2}}
\frac{\exp(15.9\,\mathrm{K}/T_\mathrm{K})}{1-\exp(-15.9\,\mathrm{K}\,T_\mathrm{K})}
\int T_\mathrm{B} \mathrm{(dv/ km\,s^{-1})},
\end{equation}

where $T_{\mathrm{K}}$ is the kinetic temperature of gas, $\sim$30\,K
(see Sect. \ref{subsectionGasTemperature}).  The brightness
temperature, $T_{\mathrm{b}}$, is estimated by
Eq. \ref{equationBrightnessTemperature} using $f_\mathrm{BEAM}$$\sim$1
and $\eta_\mathrm{MB}$$\sim$0.73.\\

Integrating Eq. \ref{equationLTE} over the area, it is possible to
derive a measurement of the mass

\begin{equation}
\label{equationLTEMass}
\ \ \ \ \ \ \ \ \ \ \ \ \ \ \ \ \ \ \ \ \ \ \ \ \ \ \ \ \ \ \ \ \ \frac{M_\mathrm{^{13}CO}}{M_\odot}
=\int m_\mathrm{H}\,\mu\,N_\mathrm{^{13}CO}\,\mathrm{d}A
\sim3.1\times10^{-11}\left[\frac{\mathrm{H_2}}{\mathrm{^{13}CO}}\right]
\frac{d}{\mathrm{kpc}}^2\alpha \frac{\exp(\frac{15.9
    \mathrm{K}}{T_\mathrm{ext}})}{1-\exp(\frac{-15.9
    \mathrm{K}}{T_\mathrm{ext}})}\int
T_\mathrm{B}\frac{\mathrm{dv}}{\mathrm{km\,s^{-1}}}\frac{\mathrm{d}\Omega}{\mathrm{arcsec}^2},
\end{equation}

where $\mu$ is the mean mass per particle, $\sim$2.29 for an H$_2$
cloud with a 25$\%$ of helium (Evans 1999), $[$H$_2$/$^{13}$CO$]$ is
the abundance ratio of H$_2$ to $^{13}$CO molecules,
$\sim$7$\times$10$^5$ \citep{frerking1982}, $\alpha$ is the H$_2$ mass
correction to include helium mass, $\sim$1.3, $\mathrm{d}A$ is the
differential area ($\mathrm{d}A=d^2\mathrm{d}\Omega$), $d$ is the
distance to the ring, $\sim$1.8\,kpc, and $\Omega$ is the solid angle.

%
%

\section{ Appendix II: Column density and mass estimated  from 1.2\,mm dust emission}

The column density can be estimated from the 1.2\,mm dust continuum
emission, $N_\mathrm{1.2mm}$.  This emission is assumed to be
optically thin, thus the total column density is given by
\citep{hildebrand1983}

\begin{equation}
\label{equationDust}
\qquad\qquad\ \ \ \ \ \ \ \ \ \ \ \ \ \ \ \ \ \ \
\ \ \ \ \ \ \ \ \ \ \ \ \ \ \ \ \ \ \
\ \ \ \ \ \ \ \ \ \ \ \ \ \ \ \ \ \ \
N_\mathrm{1.2\,mm} =
\frac{I_\mathrm{1.2\,mm}}{\mu\,m_\mathrm{H}\,k_\mathrm{1.2\,mm}\,B_\mathrm{1.2\,mm}(T_\mathrm{dust})}\frac{M_\mathrm{gas}}{M_\mathrm{dust}},
\end{equation}

where $I_\mathrm{1.2\,mm}$ is the intensity, $m_\mathrm{H}$ is the
hydrogen atom mass, $k_\mathrm{1.2\,mm}$ is the dust absorption
coefficient, $\sim$1\,cm$^2$\,g$^{-1}$ for protostellar cores
\citep{ossenkopf1994}, $B_\mathrm{1.2\,mm}(T_\mathrm{dust})$ is the
Planck function at dust temperature $T_\mathrm{dust}$, $\sim$30\,K for
regions of massive star formation \citep{faundez2004}, and
$M_\mathrm{gas}/M_\mathrm{dust}$ is the ratio of gas to dust masses,
$\sim$100 \citep{hildebrand1983}.\\

Integrating Eq. \ref{equationDust} over the area, it is possible to
derive a measurement of the mass

\begin{equation}
\label{equationDustMass}
\ \ \ \ \ \ \ \ \ \ \ \ \ \ \ \ \  \ \ \ \ \ \ \ \ \ \
\ \ \ \ \ \ \ \ \ \ \ \ \ \ \ \ \  \ \ \ \ \ \ \ \ \ \
\ \ \ \ \ \ \ \ \ 
M_\mathrm{1.2\,mm}=\int m_\mathrm{H}\,\mu\,N_\mathrm{1.2\,mm}\,\mathrm{d}A\sim
\frac{S_\mathrm{1.2\,mm}\,{d}^2}{k_\mathrm{1.2\,mm}B_\mathrm{1.2\,mm}(T_\mathrm{dust})}\frac{M_\mathrm{gas}}{M_\mathrm{dust}},
\end{equation}

where $\mathrm{d}A$ is the differential area
($\mathrm{d}A=d^2\mathrm{d}\Omega$), $d$ is the distance to the ring,
$\sim$1.8\,kpc, and $\Omega$ is the solid angle.

%
%

\section{ Appendix III: Sizes and densities}

Clump diameters, $D_\mathrm{c}$, are estimated from the deconvolved
FWHM size of their emissions, and calculated as
\begin{equation}
\label{equationSize}
\ \ \ \ \ \ \ \ \ \ \ \ \ \ \ \ \  \ \ \ \ \ \ \ \ \ \
\ \ \ \ \ \ \ \ \ \ \ \ \ \ \ \ \  \ \ \ \ \ \ \ \ \ \
\qquad\qquad\qquad\qquad
D_\mathrm{c}= d\sqrt{\theta^2_\mathrm{FWHM} - \theta^2_\mathrm{beam}},
\end{equation}
where $\theta_\mathrm{bean}$ is the beam-size, $\theta_\mathrm{FWHM}$
is the observed FWHM and $d$ is the distance to the GMC.

Given the masses and diameters of clumps, the mean column density,
N$_\mathrm{c}$, and the mean density, n$_\mathrm{c}$, are calculated
by
\begin{equation}
\label{equationColumnDensity}
\qquad\qquad\qquad\qquad 
\ \ \ \ \ \ \ \ \ \ \ \ \ \ \ \ \  \ \ \ \ \ \ \ \ \ \
\ \ \ \ \ \ \ \ \ \ \ \ \ \ \ \ \  \ \ \ \ \ \ \ \ \ \
N_\mathrm{c}= \frac{M_\mathrm{c}}{\pi(D_\mathrm{c}/2)^2} \frac{1}{\mu\,m_\mathrm{H}}
\end{equation}
and
\begin{equation}
\label{equationDensity}
\qquad\qquad\qquad\qquad 
\ \ \ \ \ \ \ \ \ \ \ \ \ \ \ \ \  \ \ \ \ \ \ \ \ \ \
\ \ \ \ \ \ \ \ \ \ \ \ \ \ \ \ \  \ \ \ \ \ \ \ \ \ \
n_\mathrm{c}= \frac{M_\mathrm{c}}{\frac{4}{3}\pi(D_\mathrm{c}/2)^3}\frac{1}{\mu\,m_\mathrm{H}},
\end{equation}
where $M_\mathrm{c}$ is the mass.

%
%

\section{ Appendix IV: Virial equilibrium}

The virial condition is 2T+W=0, where $T$ is the kinetic energy and
$W$ is the gravitational potential energy. For a clump with a
homogeneous density distribution, this condition gives

\begin{equation}
\qquad\qquad\qquad\qquad 
\ \ \ \  \ \ \ \ \ \ \ \ \ \ \ 
\ \ \ \  \ \ \ \ \ \ \ \ \ \ \ 
\ \ \ \  \ \ \ \ \ \ \ \ \ \ \ 
\ \ \ \  \ \ \ \ \ \ \ \ \ \ \ 
M_\mathrm{vir}=\frac{5\,R_\mathrm{c}\,\sigma^2_\mathrm{v}}{G},
\end{equation}

where $M_\mathrm{vir}$ is the virial mass, $R_\mathrm{c}$ is the
radius ($D_c$$/2$ ), and $\sigma_\mathrm{v}$ is the velocity
dispersion ($\sim$$\Delta V_\mathrm{c}/\sqrt{8\ln(2)}$). The parameter
$\alpha_\mathrm{virial}$ is defined as

\begin{equation}
\label{equationAlpha}
\qquad\qquad\qquad\qquad 
\ \ \ \  \ \ \ \ \ \ \ \ \ \ \ 
\ \ \ \  \ \ \ \ \ \ \ \ \ \ \ 
\ \ \ \  \ \ \ \ \ \ \ \ \ \ \ 
\ \ \ \  \ \ \ \ \ \ \ \ \ \ \ 
\ \ \ 
\alpha_\mathrm{virial} = \frac{M_\mathrm{vir}}{M_\mathrm{c}},
\end{equation}

where $M_\mathrm{c}$ is the mass of the clump.

%
%

\section{Appendix V: Bonnor-Ebert sphere}

The equation of hydrostatic equilibrium for a gas sphere is

\begin{equation}
\qquad\qquad\qquad\qquad \qquad\qquad\qquad\qquad \qquad\qquad
\ \ \ \  \ \ \ \ \  
-\frac{dp}{dr}=\frac{4{\pi}G\rho}{r^2}\int^r_0\rho\bar{r}^2d\bar{r},
\end{equation}

where $p$ is the pressure, $r$ is the distance to the center, and
$\rho$ is the density. Considering the equation of state
$p=k\,\rho\,T_\mathrm{K}/(\mu\,m_\mathrm{H})$, and making the
following substitutions $\rho=\rho_\mathrm{c}e^{-\psi}$ and
$r=\xi\,C_\mathrm{s}/\sqrt{4\,\pi\,G}$, the equation of hydrostatic
equilibrium becomes

\begin{equation}
\label{equationBonnorEbert}
\qquad\qquad\qquad\qquad \qquad\qquad\qquad\qquad \qquad\qquad
\ \ \ \  \ \ \ \ \ \ \ \ \  \ \ \ \ 
\frac{1}{\xi^2}\frac{d}{d\xi}\left(\xi^2\frac{d\psi}{\xi}\right)=e^{-\psi},
\end{equation}

where $\xi$ is a dimensionless variable, $\psi$ is a dimensionless
function, $\rho_\mathrm{c}$ is the central density (at r=0),
$C_\mathrm{s}$ is the isothermal sound speed, that is,
$C_\mathrm{s}^2=k\,T_\mathrm{k}/(\mu\,m_\mathrm{H})$ with
$T_\mathrm{K}$ as the kinetic temperature.

Imposing the boundary conditions at r=0: $\rho(r)=\rho_\mathrm{c}$ and
$d\rho(r)/dr=0$, that is, $\psi(\xi)=0$ and $d\psi(\xi)/d\xi$ at
$\xi$=0, Eq. \ref{equationBonnorEbert} is integrable numerically.  If
the gas sphere is confined by an external pressure $P_\mathrm{ext}$ at
the boundary defined by the Bonnor-Ebert radius (R$_\mathrm{B-E}$),
the solution of Eq. \ref{equationBonnorEbert} can be characterized by
the dimensionless radius $\xi_\mathrm{max}=\xi(r=R_\mathrm{B-E})$, so
$R_\mathrm{B-E}$ can be expressed as

\begin{equation}
\qquad\qquad\qquad\qquad \qquad\qquad\qquad\qquad \qquad\qquad
R_\mathrm{B-E}=\xi_\mathrm{max}\,C_\mathrm{s}/\sqrt{4\,\pi\,G\rho_\mathrm{c}},
\end{equation}
the external pressure as
\begin{equation}
\qquad\qquad\qquad\qquad  \qquad\qquad\qquad\qquad \qquad\qquad
P_\mathrm{ext}=C_\mathrm{s}^2 \left.\rho_\mathrm{c}e^{-\psi}\right|_{\xi_\mathrm{max}},
\end{equation}
and the clump mass as

\begin{equation}
\qquad\qquad\qquad\qquad \qquad\qquad\qquad\qquad \qquad\qquad
M_\mathrm{c}\sim M_\mathrm{BE}=\frac{C_\mathrm{s}^2R_\mathrm{B-E}}{G}\xi_\mathrm{max}\left.\frac{d\psi}{d\xi}\right|_{\xi_\mathrm{max}}.
\end{equation}

The stability of such a pressure-truncated gas spheres was
investigated by \cite{bonnor1956}, who showed that when
$\xi_\mathrm{max}$$>$ 6.5 the spheres are in a state of unstable
equilibrium, susceptible to gravitational collapse.  Studies of the
density profile toward isolated molecular clouds (Bok globules) in
starless stages show that $\xi_\mathrm{max}$ is concentrated near the
critical value ($\xi_\mathrm{max}$$=$ 6.5), suggesting that this
gaseous configuration represents the initial condition for star
formation in dense cores \citep{lada2007}.

Theoretically, the supporting mechanism in a Bonnor-Ebert sphere is
purely the thermal pressure, but clumps also have non-thermal support
(e.g., turbulence). The non-thermal part can be included in the model
by replacing in Eq. \ref{equationBonnorEbert} the isothermal sound
speed $C_\mathrm{s}$ by an effective speed $C_\mathrm{eff}$, where

\begin{equation}
\qquad\qquad\qquad\qquad  \qquad\qquad\qquad\qquad \qquad\qquad
C_\mathrm{eff}^2 = C_\mathrm{s}^2 + C_\mathrm{NT}^2,
\end{equation}

with $C_\mathrm{NT}$ as the non-thermal speed
\citep[e.g.][]{kandori2005}.  The effective speed is approximated as

\begin{equation}
\qquad\qquad\qquad\qquad  \qquad\qquad\qquad\qquad \qquad\qquad
C_\mathrm{eff}\sim\frac{\Delta V_\mathrm{c}}{\sqrt{8\ln(2)}},
\end{equation}
and the Bonnor-Ebert radius as 
\begin{equation}
\qquad\qquad\qquad\qquad  \qquad\qquad\qquad\qquad \qquad\qquad
R_\mathrm{B-E}\sim\frac{D_\mathrm{c}}{2},
\end{equation}

where $D_\mathrm{c}$ and $\Delta V_\mathrm{c}$ are the diameter and
line velocity width (FWHM) of the clumps, respectively.

\clearpage 
\onecolumn

\setlongtables
\setlength\LTcapwidth{480pt}

\onltab{7}{
\newcommand{\tableA}{Name & R.A. & Dec.& $D_\mathrm{c}$  & $M_\mathrm{c}$ & $\Delta V_\mathrm{c}$& $n_\mathrm{c}$ & $N_\mathrm{c}$ & Clump Type & B-E \\     
&\multicolumn{2}{c}{(J2000)}  &pc        &  M$_\odot$ & km\,s$^{-1}$ & cm$^{-3}$ & cm$^{-2}$  & &\\
}
\begin{center}

\begin{longtable}{cccccccccc}
\caption{Physical properties of the $^{13}$CO clumps. Column 1 gives names; columns 2 and 3, equatorial coordinates;
column 4, diameters; column 5, masses; column 6, line velocity widths; column 7, densities; column 8, column densities;
column 9, clump types;
and column 10, if clumps are candidates to be experimenting collapse ($\xi_\mathrm{max}$ $\ge$ 6.5).
}\\
\label{table13COClumps}\\
\endfirsthead
\multicolumn{4}{c}%
{{\bfseries \tablename\thetable{} -- continued from previous page}} \\
\hline\hline\\
\tableA
\hline\\
\endhead
\hline\\
\multicolumn{4}{c}{{Continued on next page}}\\
\endfoot
\hline\\
\endlastfoot
\hline\hline\\
\tableA
\hline\\
1 & 16:59:06.92 & -40:13:38.37 & 1.0 &2.3x10$^{2}$  & 2.0 &8.7x10$^{3}$  &1.7x10$^{22}$ & A& Y
 \\ 
2 & 16:59:41.33 & -40:03:46.92 & 1.6 &7.5x10$^{2}$  & 4.2 &6.3x10$^{3}$  &2.1x10$^{22}$ & A& N
 \\ 
3 & 16:59:45.58 & -40:10:16.90 & 1.0 &1.3x10$^{2}$  & 1.3 &4.1x10$^{3}$  &8.6x10$^{21}$ & C& Y
 \\ 
4 & 16:59:48.16 & -40:10:54.80 & 1.3 &2.4x10$^{2}$  & 1.2 &3.4x10$^{3}$  &9.5x10$^{21}$ & C& Y
 \\ 
5 & 16:59:07.18 & -40:06:05.01 & 0.9 &1.3x10$^{2}$  & 1.8 &6.7x10$^{3}$  &1.2x10$^{22}$ & B& N
 \\ 
6 & 16:59:11.91 & -40:11:05.78 & 0.7 &4.5x10$^{1}$  & 1.1 &5.2x10$^{3}$  &7.1x10$^{21}$ & B& Y
 \\ 
7 & 16:59:21.32 & -40:11:08.42 & 1.0 &1.2x10$^{2}$  & 1.5 &4.4x10$^{3}$  &8.9x10$^{21}$ & B& Y
 \\ 
8 & 16:59:08.19 & -40:12:54.83 & 1.0 &8.0x10$^{1}$  & 1.4 &2.7x10$^{3}$  &5.5x10$^{21}$ & C& N
 \\ 
9 & 16:59:49.55 & -40:11:46.04 & 0.9 &7.7x10$^{1}$  & 1.1 &4.1x10$^{3}$  &7.2x10$^{21}$ & C& Y
 \\ 
10 & 16:59:18.73 & -40:10:59.94 & 1.0 &4.0x10$^{1}$  & 0.8 &1.4x10$^{3}$  &2.8x10$^{21}$ & A& Y
 \\ 
11 & 16:59:29.53 & -40:10:14.71 & 1.2 &1.3x10$^{2}$  & 1.4 &2.4x10$^{3}$  &6.0x10$^{21}$ & B& Y
 \\ 
12 & 16:59:34.55 & -40:07:36.79 & 0.8 &4.0x10$^{1}$  & 1.2 &2.2x10$^{3}$  &3.9x10$^{21}$ & B& N
 \\ 
13 & 16:59:04.56 & -40:11:24.80 & 0.9 &3.1x10$^{1}$  & 1.0 &1.6x10$^{3}$  &2.8x10$^{21}$ & B& N
 \\ 
14 & 16:59:25.25 & -40:12:52.81 & 1.1 &4.2x10$^{1}$  & 1.2 &1.2x10$^{3}$  &2.5x10$^{21}$ & C& N
 \\ 
15 & 16:59:15.70 & -40:10:00.09 & 1.0 &9.0x10$^{1}$  & 1.3 &2.7x10$^{3}$  &5.7x10$^{21}$ & C& Y
 \\ 
16 & 16:59:36.19 & -40:12:56.33 & 0.4 &3.1x10$^{1}$  & 2.6 &1.4x10$^{4}$  &1.2x10$^{22}$ & B& N
 \\ 
17 & 16:59:34.67 & -40:09:59.05 & 1.4 &1.5x10$^{2}$  & 1.3 &1.8x10$^{3}$  &5.2x10$^{21}$ & B& Y
 \\ 
18 & 16:59:53.37 & -40:09:04.01 & 1.1 &1.3x10$^{2}$  & 1.9 &2.9x10$^{3}$  &6.8x10$^{21}$ & B& N
 \\ 
19 & 16:59:37.24 & -40:08:32.24 & 0.9 &4.2x10$^{1}$  & 0.7 &1.8x10$^{3}$  &3.4x10$^{21}$ & C& Y
 \\ 
20 & 16:59:43.02 & -40:08:11.25 & 1.4 &1.3x10$^{2}$  & 1.4 &1.6x10$^{3}$  &4.6x10$^{21}$ & B& N
 \\ 
21 & 16:59:46.10 & -40:09:20.88 & 1.1 &7.3x10$^{1}$  & 1.4 &2.0x10$^{3}$  &4.4x10$^{21}$ & C& N
 \\ 
22 & 16:59:22.89 & -39:58:07.42 & 0.8 &5.0x10$^{1}$  & 1.6 &2.9x10$^{3}$  &5.0x10$^{21}$ & C& N
 \\ 
23 & 16:59:02.43 & -40:04:56.57 & 1.1 &4.4x10$^{1}$  & 1.6 &1.2x10$^{3}$  &2.6x10$^{21}$ & C& N
 \\ 
24 & 16:59:12.32 & -40:09:06.94 & 0.8 &2.8x10$^{1}$  & 1.1 &1.9x10$^{3}$  &3.1x10$^{21}$ & C& N
 \\ 
25 & 16:59:47.41 & -40:02:32.38 & 1.0 &6.4x10$^{1}$  & 1.7 &2.0x10$^{3}$  &4.3x10$^{21}$ & C& N
 \\ 
26 & 16:59:50.33 & -40:00:12.50 & 1.1 &4.7x10$^{1}$  & 1.0 &1.3x10$^{3}$  &2.8x10$^{21}$ & A& N
 \\ 
27 & 16:59:15.12 & -40:00:50.15 & 0.6 &1.2x10$^{1}$  & 1.2 &1.6x10$^{3}$  &2.1x10$^{21}$ & B& N
 \\ 
28 & 16:59:08.70 & -40:02:52.08 & 0.7 &3.6x10$^{1}$  & 2.0 &3.2x10$^{3}$  &4.8x10$^{21}$ & B& N
 \\ 
29 & 16:59:30.75 & -39:57:57.01 & 0.8 &2.9x10$^{1}$  & 0.8 &1.8x10$^{3}$  &3.0x10$^{21}$ & C& N
 \\ 
30 & 16:59:41.06 & -39:58:24.31 & 0.9 &4.4x10$^{1}$  & 1.2 &2.1x10$^{3}$  &3.9x10$^{21}$ & B& N
 \\ 
31 & 16:59:42.13 & -40:10:51.31 & 1.2 &3.8x10$^{1}$  & 1.1 &8.5x10$^{2}$  &2.0x10$^{21}$ & B& N
 \\ 
32 & 16:59:13.96 & -40:13:17.77 & 1.2 &5.7x10$^{1}$  & 1.4 &1.2x10$^{3}$  &3.0x10$^{21}$ & C& N
 \\ 
33 & 16:59:46.49 & -40:02:09.44 & 1.5 &1.3x10$^{2}$  & 2.1 &1.4x10$^{3}$  &4.3x10$^{21}$ & C& N
 \\ 
34 & 16:59:13.72 & -40:14:22.98 & 0.8 &2.9x10$^{1}$  & 1.2 &1.7x10$^{3}$  &2.9x10$^{21}$ & C& N
 \\ 
35 & 16:59:17.95 & -40:11:17.31 & 1.3 &5.7x10$^{1}$  & 1.2 &8.6x10$^{2}$  &2.3x10$^{21}$ & A& N
 \\ 
36 & 16:59:48.36 & -40:08:37.90 & 0.8 &2.2x10$^{1}$  & 1.5 &1.6x10$^{3}$  &2.5x10$^{21}$ & C& N
 \\ 
37 & 16:59:33.77 & -39:56:56.65 & 0.8 &2.2x10$^{1}$  & 0.9 &1.5x10$^{3}$  &2.4x10$^{21}$ & C& N
 \\ 
38 & 16:59:09.04 & -40:10:23.56 & 1.1 &3.1x10$^{1}$  & 1.2 &8.9x10$^{2}$  &1.9x10$^{21}$ & C& N
 \\ 
39 & 16:59:05.79 & -40:06:35.22 & 1.2 &3.4x10$^{1}$  & 1.4 &7.2x10$^{2}$  &1.7x10$^{21}$ & B& N
 \\ 
40 & 16:59:28.09 & -39:58:06.82 & 0.9 &2.1x10$^{1}$  & 0.6 &1.0x10$^{3}$  &1.9x10$^{21}$ & B& Y
 \\ 
41 & 16:59:17.31 & -39:58:25.14 & 0.8 &1.6x10$^{1}$  & 1.4 &1.1x10$^{3}$  &1.8x10$^{21}$ & C& N
 \\ 
42 & 16:59:44.41 & -39:59:30.08 & 0.8 &2.5x10$^{1}$  & 1.2 &1.7x10$^{3}$  &2.8x10$^{21}$ & B& N
 \\ 
43 & 16:58:57.39 & -40:05:28.90 & 1.1 &3.0x10$^{1}$  & 1.5 &8.3x10$^{2}$  &1.8x10$^{21}$ & C& N
 \\ 
44 & 16:59:31.42 & -40:06:57.92 & 0.9 &5.3x10$^{1}$  & 1.3 &2.2x10$^{3}$  &4.2x10$^{21}$ & B& N
 \\ 
45 & 16:59:29.00 & -39:56:30.82 & 0.8 &1.5x10$^{1}$  & 0.6 &9.0x10$^{2}$  &1.5x10$^{21}$ & C& N
 \\ 
46 & 16:59:25.85 & -40:13:17.45 & 0.7 &1.2x10$^{1}$  & 1.3 &1.1x10$^{3}$  &1.6x10$^{21}$ & C& N
 \\ 
47 & 16:59:47.47 & -39:59:33.29 & 1.1 &4.8x10$^{1}$  & 1.0 &1.3x10$^{3}$  &2.8x10$^{21}$ & C& N
 \\ 
48 & 16:59:12.24 & -40:02:06.95 & 0.5 &1.1x10$^{1}$  & 0.9 &2.7x10$^{3}$  &2.9x10$^{21}$ & B& N
 \\ 
49 & 16:59:54.30 & -40:13:01.26 & 0.7 &1.6x10$^{1}$  & 1.6 &1.8x10$^{3}$  &2.5x10$^{21}$ & C& N
 \\ 
50 & 16:59:08.43 & -40:09:49.16 & 0.8 &1.3x10$^{1}$  & 0.9 &8.9x10$^{2}$  &1.5x10$^{21}$ & C& N
 \\ 
51 & 16:58:48.70 & -40:08:06.74 & 0.6 &1.5x10$^{1}$  & 1.0 &2.0x10$^{3}$  &2.5x10$^{21}$ & C& N
 \\ 
52 & 16:58:55.54 & -40:08:51.93 & 0.6 &6.5x10$^{0}$  & 0.9 &1.1x10$^{3}$  &1.3x10$^{21}$ & C& N
 \\ 
53 & 16:59:21.67 & -40:09:28.99 & 0.9 &2.9x10$^{1}$  & 1.0 &1.4x10$^{3}$  &2.6x10$^{21}$ & C& N
 \\ 
54 & 16:59:50.98 & -40:03:46.26 & 1.0 &2.9x10$^{1}$  & 1.8 &1.1x10$^{3}$  &2.2x10$^{21}$ & C& N
 \\ 
55 & 17:00:00.54 & -40:09:38.67 & 1.1 &6.8x10$^{1}$  & 1.7 &1.8x10$^{3}$  &4.0x10$^{21}$ & B& N
 \\ 
56 & 16:59:09.15 & -40:01:49.37 & 0.7 &8.1x10$^{0}$  & 0.6 &1.0x10$^{3}$  &1.3x10$^{21}$ & C& N
 \\ 
57 & 16:59:37.21 & -40:07:19.98 & 1.3 &2.6x10$^{1}$  & 1.0 &4.4x10$^{2}$  &1.1x10$^{21}$ & B& N
 \\ 
58 & 16:59:48.75 & -40:05:13.00 & 1.3 &2.8x10$^{1}$  & 1.1 &4.3x10$^{2}$  &1.1x10$^{21}$ & C& N
 \\ 
59 & 16:59:05.32 & -39:56:06.84 & 0.5 &1.4x10$^{1}$  & 1.9 &4.2x10$^{3}$  &4.1x10$^{21}$ & B& N
 \\ 
60 & 16:59:58.09 & -40:08:07.26 & 1.0 &3.4x10$^{1}$  & 1.2 &1.0x10$^{3}$  &2.2x10$^{21}$ & C& N
 \\ 
61 & 16:59:00.16 & -40:10:57.29 & 0.7 &9.2x10$^{0}$  & 0.7 &7.6x10$^{2}$  &1.2x10$^{21}$ & C& N
 \\ 
62 & 16:59:13.59 & -40:10:51.76 & 1.3 &1.5x10$^{1}$  & 1.0 &2.2x10$^{2}$  &6.0x10$^{20}$ & B& N
 \\ 
63 & 16:59:44.79 & -40:07:57.18 & 1.1 &3.3x10$^{1}$  & 0.8 &9.4x10$^{2}$  &2.1x10$^{21}$ & C& Y
 \\ 
64 & 16:59:41.38 & -39:59:31.70 & 1.1 &2.3x10$^{1}$  & 1.1 &6.0x10$^{2}$  &1.3x10$^{21}$ & B& N
 \\ 
65 & 16:59:02.30 & -40:09:45.35 & 0.7 &4.1x10$^{0}$  & 0.7 &4.4x10$^{2}$  &6.1x10$^{20}$ & C& N
 \\ 
66 & 16:59:31.41 & -40:04:52.24 & 0.8 &1.6x10$^{1}$  & 1.7 &1.1x10$^{3}$  &1.8x10$^{21}$ & C& N
 \\ 
67 & 16:59:05.65 & -39:56:18.12 & 0.8 &1.9x10$^{1}$  & 1.3 &1.2x10$^{3}$  &2.0x10$^{21}$ & B& N
 \\ 
68 & 16:59:29.06 & -39:56:49.57 & 0.8 &8.4x10$^{0}$  & 0.8 &6.0x10$^{2}$  &9.6x10$^{20}$ & C& N
 \\ 
69 & 16:59:45.98 & -39:57:07.56 & 1.0 &2.5x10$^{1}$  & 0.8 &8.8x10$^{2}$  &1.8x10$^{21}$ & C& N
 \\ 
70 & 16:58:51.12 & -40:09:22.29 & 0.7 &1.1x10$^{1}$  & 1.3 &9.6x10$^{2}$  &1.5x10$^{21}$ & C& N
 \\ 
71 & 16:59:49.75 & -40:12:19.65 & 1.0 &1.3x10$^{1}$  & 0.9 &4.2x10$^{2}$  &8.8x10$^{20}$ & C& N
 \\ 
72 & 17:00:01.15 & -40:10:36.04 & 1.2 &8.4x10$^{0}$  & 0.8 &1.8x10$^{2}$  &4.3x10$^{20}$ & C& N
 \\ 
73 & 16:59:25.56 & -40:14:09.62 & 0.8 &3.4x10$^{0}$  & 0.4 &2.4x10$^{2}$  &3.8x10$^{20}$ & C& N
 \\ 
74 & 17:00:04.98 & -40:13:51.41 & 0.6 &3.0x10$^{0}$  & 0.8 &4.0x10$^{2}$  &5.2x10$^{20}$ & C& N
 \\ 
75 & 16:59:53.42 & -40:07:18.93 & 1.1 &3.1x10$^{1}$  & 1.1 &7.8x10$^{2}$  &1.8x10$^{21}$ & C& N
 \\ 
76 & 16:59:50.12 & -39:56:08.24 & 0.6 &8.0x10$^{0}$  & 0.6 &1.1x10$^{3}$  &1.4x10$^{21}$ & C& N
 \\ 
77 & 16:59:54.21 & -39:56:12.17 & 0.6 &5.4x10$^{0}$  & 0.7 &7.3x10$^{2}$  &9.5x10$^{20}$ & C& N
 \\ 
78 & 16:58:56.16 & -40:10:17.33 & 0.6 &4.8x10$^{0}$  & 0.7 &7.2x10$^{2}$  &9.0x10$^{20}$ & C& N
 \\ 
79 & 16:59:55.60 & -39:58:21.95 & 0.9 &2.3x10$^{1}$  & 0.9 &1.0x10$^{3}$  &1.9x10$^{21}$ & B& N
 \\ 
80 & 16:58:54.33 & -40:10:02.60 & 0.4 &3.0x10$^{0}$  & 0.8 &2.4x10$^{3}$  &1.7x10$^{21}$ & C& N
 \\ 
81 & 17:00:12.15 & -40:11:00.99 & 0.7 &4.7x10$^{0}$  & 1.5 &4.2x10$^{2}$  &6.3x10$^{20}$ & C& N
 \\ 
82 & 16:59:59.63 & -40:06:11.79 & 0.9 &9.5x10$^{0}$  & 1.5 &4.2x10$^{2}$  &7.9x10$^{20}$ & C& N
 \\ 
83 & 16:59:55.18 & -39:58:56.77 & 0.9 &1.7x10$^{1}$  & 0.7 &7.4x10$^{2}$  &1.4x10$^{21}$ & B& N
 \\ 
84 & 16:59:25.33 & -40:08:16.73 & 0.7 &1.1x10$^{1}$  & 0.9 &1.3x10$^{3}$  &1.7x10$^{21}$ & C& N
 \\ 
85 & 16:59:31.92 & -40:09:48.39 & 1.2 &1.4x10$^{1}$  & 0.9 &2.6x10$^{2}$  &6.4x10$^{20}$ & C& N
 \\ 
86 & 16:59:56.61 & -39:56:36.53 & 0.7 &7.9x10$^{0}$  & 0.9 &9.7x10$^{2}$  &1.3x10$^{21}$ & C& N
 \\ 
87 & 17:00:04.92 & -39:57:44.23 & 0.4 &4.5x10$^{0}$  & 0.7 &2.8x10$^{3}$  &2.2x10$^{21}$ & C& N
 \\ 
88 & 16:59:54.33 & -40:00:52.24 & 1.0 &1.4x10$^{1}$  & 0.8 &5.1x10$^{2}$  &1.0x10$^{21}$ & C& N
 \\ 
89 & 16:59:41.43 & -39:56:45.98 & 0.8 &1.5x10$^{1}$  & 0.9 &9.3x10$^{2}$  &1.6x10$^{21}$ & C& N
 \\ 
90 & 16:59:26.35 & -40:14:22.66 & 0.6 &2.9x10$^{0}$  & 0.6 &4.3x10$^{2}$  &5.4x10$^{20}$ & C& N
 \\ 
91 & 16:59:58.63 & -39:55:16.58 & 0.7 &8.5x10$^{0}$  & 0.7 &7.1x10$^{2}$  &1.1x10$^{21}$ & C& N
 \\ 
92 & 16:59:43.34 & -40:13:25.33 & 0.3 &2.3x10$^{0}$  & 0.5 &2.9x10$^{3}$  &1.8x10$^{21}$ & C& N
 \\ 
93 & 16:59:58.94 & -39:57:00.07 & 0.6 &3.7x10$^{0}$  & 0.6 &5.4x10$^{2}$  &6.8x10$^{20}$ & C& N
 \\ 
94 & 16:59:43.79 & -40:06:24.61 & 0.9 &1.8x10$^{1}$  & 1.2 &7.5x10$^{2}$  &1.4x10$^{21}$ & A& N
 \\ 
95 & 16:59:46.42 & -39:58:02.25 & 0.8 &1.3x10$^{1}$  & 0.9 &7.7x10$^{2}$  &1.3x10$^{21}$ & C& N
 \\ 
96 & 16:59:51.38 & -39:55:57.85 & 0.6 &6.4x10$^{0}$  & 0.7 &9.5x10$^{2}$  &1.2x10$^{21}$ & C& N
 \\ 
97 & 17:00:01.61 & -40:10:06.77 & 1.5 &5.1x10$^{1}$  & 1.3 &5.6x10$^{2}$  &1.7x10$^{21}$ & B& N
 \\ 
98 & 17:00:01.71 & -39:57:22.28 & 0.7 &4.4x10$^{0}$  & 0.7 &4.7x10$^{2}$  &6.5x10$^{20}$ & C& N
 \\ 
99 & 16:59:41.32 & -39:57:09.75 & 1.0 &1.1x10$^{1}$  & 1.0 &4.0x10$^{2}$  &8.1x10$^{20}$ & C& N
 \\ 
100 & 17:00:01.60 & -40:05:21.98 & 0.9 &1.2x10$^{1}$  & 0.9 &5.5x10$^{2}$  &1.0x10$^{21}$ & C& N
 \\ 
101 & 16:59:33.24 & -40:13:18.61 & 1.3 &6.7x10$^{0}$  & 0.9 &1.0x10$^{2}$  &2.7x10$^{20}$ & C& N
 \\ 
102 & 17:00:09.38 & -39:57:04.29 & 0.5 &2.7x10$^{0}$  & 0.7 &8.8x10$^{2}$  &8.5x10$^{20}$ & C& N
 \\ 
103 & 16:59:54.25 & -40:10:14.04 & 0.9 &5.7x10$^{0}$  & 1.1 &2.9x10$^{2}$  &5.2x10$^{20}$ & C& N
 \\ 
104 & 16:59:47.17 & -40:12:36.94 & 0.7 &4.9x10$^{0}$  & 1.0 &4.3x10$^{2}$  &6.4x10$^{20}$ & C& N
 \\ 

\end{longtable}
\end{center}
}

\ifshowRecord

\newpage
\section*{Change Record}

\begin{table}[htbp]
\centerline{
\begin{tabular}{|c|c|c|c|p{200pt}|}
  \hline
 \rowcolor{gray}  Version   &   Date  &  Affected Section(s)    &  Author &  Reason, Initiation, Remarks\\\hline
 V6     &  2013-01-25  & All     &  & Per's corrections. Lars-{\AA}ke's corrections. Included outflow 2.Included two filaments.\\
\hline
 V7     &  2013-03-06  & All     &  & Itziar's corrections.\\
\hline
 V8     &  2013-04-04  & All     &  & New fit for the expanding ring. Per's  corrections.\\
\hline
V9     &  2013-04-08  & All     &  & Per's corrections.\\
\hline
V10     &  2013-04-13  & All     &  & Lars-{\AA}ke's corrections. New fit for the expanding ring.\\
\hline
V11     &  2013-04-14  & All     &  & Submitted.\\
\hline
V12     &  2013-04-14  & All     &  & Leo's corrections.\\
\hline
\end{tabular}}

\end{table}

\fi

\end{document}